\documentclass[12pt]{article}
\pdfoutput=1
\usepackage[utf8]{inputenc}
\usepackage{bm}
\usepackage{mathrsfs}
\usepackage{empheq}
\usepackage{verbatim}
\usepackage{bibentry}
\usepackage{jheppub}
\usepackage{hyperref}
\usepackage{slashed}
\usepackage{subcaption}
\usepackage{multirow}
\usepackage{color}
\usepackage{tikz}
\usepackage{array}
\usepackage{tablefootnote}
\usepackage{latexsym}
\usepackage{amssymb}
\usepackage{amsmath}
\usepackage{mathtools}
\usepackage{stackrel}
\usepackage{youngtab}
\usepackage{cancel}
\usepackage[nosort]{cite}
\usepackage{graphicx}
\usepackage{epsfig}
\usepackage{psfrag}
\usepackage{tensor}
\usepackage{ulem}

\usepackage{tocloft} 
\usepackage{ulem}
\usepackage{soul,xcolor}
\setstcolor{red}

\newcommand{\nin} {\noindent}
\newcommand{\nn}{\nonumber}
\newcommand{\be}{\begin{equation}}
\newcommand{\ee}{\end{equation}}

\numberwithin{equation}{section}

\title{Notes on Gauge Fields and Discrete Series representations in de Sitter spacetimes}

\author[1]{Alan Rios Fukelman,}
\author[2]{Mat\'\i as Semp\'e}
\author[2,3]{and Guillermo A. Silva}

\affiliation[1]{Department of Mathematics, King's College London \\ The Strand, London WC2R 2LS, U.K.}
\affiliation[2]{Instituto de F\'isica de La Plata - CONICET\\
Diagonal 113 e/ 63 y 64, 1900 - La Plata, Argentina}
\affiliation[3]{Departamento de F\'isica, Universidad Nacional de La Plata \\ C.C. 67, 1900 - La Plata, Argentina}

\emailAdd{alan.rios\_fukelman@kcl.ac.uk}
\emailAdd{silva@fisica.unlp.edu.ar}
\emailAdd{sempe\_100@hotmail.com}

\abstract{In this note we discuss features of the simplest spinning Discrete Series Unitary Irreducible Representations (UIR) of SO(1,4). These representations are known to be realised in the single particle Hilbert space of a free gauge field propagating in a four dimensional fixed de Sitter background. They showcase distinct features as compared to the more common Principal Series realised by heavy  fields. Upon computing the $1-$loop Sphere path integral we show that the \emph{edge modes} of the theory can be understood in terms of a Discrete Series of SO$(1,2)$. We then canonically quantise the theory and show how group theory constrains the mode decomposition. We further clarify the role played by the second SO(4) Casimir in the single particle Hilbert space of the theory.
}

\begin{document}
\maketitle

\section{Motivations}

de Sitter spacetime (dS) is the simplest example of an accelerating cosmology such as the one our universe is evolving to \citep{planck,Perlmutter_1999,baryonacust}. Despite this, finding the right theoretical framework to understand its quantum features has proven to be a challenging problem. The lack of an S-matrix or a boundary in which to anchor an observer makes the definition of physical observables, and even computables, in dS a subtle issue. Nevertheless, there has still been steady progress in the field.

Building on recent ideas developed in the context of the AdS/CFT duality, there has been a new influx of developments that try to overcome some of these shortcomings. Realising a portion of de Sitter spacetime within the deep interior of a $(d+1)-$dimensional anti-de Sitter gives a natural spatial boundary in which to anchor an observer and even more it might give the framework in which to try and define a dual holographic model. This idea was realised for the case of an AdS$_2$ geometry in \citep{Anninos:2017hhn,Anninos:2018svg}. Microscopically, a flow between two different AdS$_2$ regions can be understood by a RG-flow of a deformed SYK model. This also opens up the possibility of realising the portion of dS$_2$ inside the AdS$_2$ geometry microscopically by a specific deformation of such RG-flows \citep{Anninos:2020cwo,Anninos:2022qgy}. Further ideas along these lines both for SYK models and JT gravity can be found in \citep{Susskind:2021esx,Ecker:2022vkr,Anninos:2022hqo,Witten:2020ert}. For $D = 2+1$, a generalisation of the $T\bar{T}$ deformation \citep{Gorbenko:2018oov,Lewkowycz:2019xse} allows one to reconstruct patches of AdS or dS from a deformed holographic seed CFT formulated at the boundary of a patch \citep{Coleman:2021nor,Shyam:2021ciy,Dong:2018cuv}. These set of ideas provide setups to understand   de Sitter microstates as a re-organisation of AdS ones.
 
\subsection*{Macroscopic Considerations}

\nin While we develop a complete microscopic formulation of dS spacetime, much can be learned from studying the representation theory of the SO$(1,d+1)$ isometry group of dS$_{d+1}.$\footnote{As well known, SO$(1,d+1$)  is also the $d-$dimensional Euclidean conformal group.} Since the foundational work of Wigner in Minkowski spacetime \citep{wigner} we known that Unitary Irreducible Representations (UIRs)  of the Poincare group are related  to elementary particles represented as free quantum fields. The UIR labels in 4d Minkowski space are $(m^2,s)$, these are, respectively, the mass and spin of the corresponding field. Similar ideas have been applied to the SO$(2,d)$ isometry group of $(d+1)-$dimensional anti-de Sitter  spacetimes (AdS$_{d+1}$) \citep{Minwalla:1997ka, deWit:1999ui,Dolan:2005wy, Rychkov:2016iqz}. 
In this case the UIRs labels  $(\Delta,\bm s)$ are associated to the maximal compact subgroup SO$(2)\,\times\,$SO($d)\subset$ SO$(2,d)$, they are called scaling dimension and spin respectively. For the AdS case, the SO(2) quantum number  corresponds to the charge associated to the globally defined time-like Killing vector $\bm\partial_t$ in AdS global coordinates. Thus,  on general grounds, the relevant UIRs  in AdS are those for which the spectrum of  $\mathfrak{so}(2)$ is bounded.

The  de Sitter spacetime case is more involved. To start, we remark that states in a dS multiplet cannot, in general, be classified using the maximally compact subgroup SO$(d+1)\subset$  SO$(1,d+1$). Indeed, in odd spacetime dimensions  the number of Cartan's in SO$(d+1)$ are not enough.  Instead,  states are classified using SO$(1,1)\,\times\,$ SO$(d)\subset$ SO$(1,d+1)$ \citep{Dobrev:1977qv,Sun:2021thf} with  labels denoted in the same way as in AdS, i.e. $(\Delta, \bm s)$ respectively. However, an important difference as compared to the anti-de Sitter case is that the label $\Delta$ now corresponds to the non-compact generator of $\mathfrak{so}(1,1)$. As a result, complex values for $\Delta$ show up in the classification of UIRs.  Intriguingly, the  $\mathfrak{so}(1,1)$ quantum number relates to static patch time coordinate. It is important to remark that de Sitter spacetime does not have a globally defined timelike Killing vector. This precludes any global definition of conserved energy in the spacetime.

\subsection*{SO$(1,d+1)$ Basics}

\nin SO$(1,d+1)$ admits UIRs with  complex scaling dimension \citep{Dobrev:1977qv}. Although perhaps a perplexing fact, the simplest case of  a scalar field with mass-squared above the de Sitter scale  turns out to correspond to such complex $\Delta$.\footnote{Massive fermionic fields also have a complex scaling dimension $\Delta$ (see \citep{Schaub:2023scu} for recent work).} 
Hence, the single particle Hilbert space of heavy scalar fields realises what is known as the Principal Series representation. 
These representations are seen to appear in numerous physical interesting scenarios (see \citep{Anous:2020nxu} and references therein), and they also show up for fermionic and massive spin$-s$ fields \citep{Sun:2021thf,Basile:2018zoy,Letsios:2020twa,Letsios:2022slc,Pethybridge:2021rwf, Schaub:2023scu}. However, we stress that  they do not account for the more interesting cases of gauge fields in de Sitter.  

As first noted for the massive spin$-2$ field in \citep{Deser:1983mm,Higuchi:1986py}, and further studied for higher spin fields  in \citep{Deser:2001pe}, specific tunings of the mass parameter and the cosmological constant furnish unitary single particle Hilbert spaces with reduced degrees of freedom. This is achieved by the appearance of gauge invariance in the equations of motion. From a representation theory point of view these field equations were shown to generically correspond to UIR of SO$(1,d+1)$ known as Exceptional Series. In particular, in even spacetime dimensions gauge fields either belong to the Exceptional or Discrete Series representation. The characteristic  feature of Discrete Series is that their corresponding weight vectors have no vanishing entries, whereas Exceptional Series UIRs weight vectors have at least one zero component (see \citep{Basile:2016aen} for further details). For lower even dimensions, $D = 2, 4$, weight vectors have only one component, hence, spinning Exceptional and Discrete UIRs coincide. 
From now on we will abuse language and refer to maximally symmetric traceless tensors generically as Discrete UIRs, since we will be working in $D=4$.\footnote{A relevant point to keep in mind is that a theorem of Harish-Chandra establishes that a real Semisimple Lie group possesses Discrete Series UIRs if and only if it has a compact Cartan subgroup. Hence Discrete UIRs only show up in even dimensional de Sitter spacetimes (thm. 12.20 in \citep{knapp}.).} The characteristic feature of Exceptional/Discrete series is that $2\Delta \in \mathbb{Z}$.

The simplest situation where Discrete representations arise is for SO$(1,2)$ which happens to be the isometry group of dS$_2$ (see \citep{Bargmann:1946me,harishplancharel,thomas}). Perhaps surprisingly, a field theoretic realisation devoid of problems has been elusive till recently. They were shown to arise in arbitrary dimensions in the single-particle Hilbert space of free tachyonic scalar models \citep{Bros:2010wa, Epstein:2014jaa}. Even the simplest case corresponding to a massless scalar field suffers from pathologies that can be traced back to a problematic zero mode \citep{Anninos:2023lin}. One could then be lead to believe that, at least in $2-$dimensions, these representations shouldn't be taken seriously. However, further considerations have  shown that this is not the case as discrete series UIRs appear generally in the multiparticle (tensor-product) Hilbert space of Principal Series fields \citep{Dobrev:1977qv,repka}. Stated otherwise, Discrete UIR  particles can be generated through the decay of two heavy fields in an interacting QFT  \citep{Nachtmann1968DynamischeSI,Bros:2010wa}.
  
Discrete UIRs can also show up for the case of fermionic fields. The simplest example being $\Delta = \frac{3}{2}$ corresponding to a 2d gravitino. A local action for such a field requires a pure imaginary \emph{mass} term. This signals some tension between the field theory, in which a complex action would render the theory non-hermitian and hence not unitary, and Group Theory, which demands an  imaginary mass to have a UIR. In 2d this tension is alleviated since the canonical kinetic term for the gravitino exactly vanishes,\footnote{In 2d there is no rank$-3$ Clifford algebra element $\gamma^{\mu \nu \rho} \equiv \gamma^{[\mu} \gamma^\nu \gamma^{\rho ]} = 0$} leading to gravitini with no propagating degrees of freedom \citep{Anninos:2023exn}. Furthermore, even with imaginary \emph{mass}, upon properly treating the gauge invariance, the 2d gravitino renders a sensible Euclidean path integral that yields at one loop the character of the corresponding UIR \citep{Anninos:2023exn}.

In four dimensional dS$_4$ the situation is even more interesting, as  Discrete Series UIR appears in the single-particle Hilbert space of massless and partially massless $s\ge1$ higher-spin gauge fields. In particular, Discrete series describe the Hilbert space of the spin$-2$ field corresponding to the graviton \citep{Higuchi:1991tn}.\footnote{The 4d gravitino also showcases similar features to the 2d Discrete Series  case.}
There is still one caveat, fields realising the Discrete Series representation involve an underlying gauge invariance which has to be properly gauge fixed to render a sensible theory. As recently shown for the massless scalar field 2d \citep{Anninos:2023lin}, proper account of the shift  symmetry removes the problematic zero at the expense of the inclusion of new gauge fields. The final result agrees with expectations rendering the single particle Hilbert space of the theory equivalent to the Discrete Series representation. One of the motivations of the present work is to extend this approach to higher dimensions. Analysis of fermionic fields in de Sitter space have been recently considered in \citep{Letsios:2020twa,Letsios:2022slc,Letsios:2023qzq}. %
\\ 

\nin \textbf{An Euclidean Perspective}
\vspace{2mm}

\nin Even after furnishing the corresponding  UIR field theoretically, the basic problem of finding a set of sharp observables still persists. There is a calculable quantity at our disposal that has all the symmetries gauged, it is  gauge and field redefinition invariant and contains physically relevant information, namely the Euclidean path integral. Following insights from black hole physics \citep{Bekenstein:1972tm,Hawking:1975vcx}, Gibbons and Hawking \citep{Gibbons:1977mu} proposed that the area of the dS cosmological horizon should be interpreted as an entropy.\footnote{Recall that a free falling observer following a timelike geodesic in a fixed dS background is only causally connected to a region of spacetime bounded by past and future horizons, namely the cosmological horizon.} This entropy is then macroscopically defined by the Euclidean path integral\footnote{On general grounds we  identify Euclidean path integrals with the Helmholtz free energy of the system, ${\cal Z}=e^{-\beta F}$ where $F=E-T S$ with $E$ the energy, $T$ the temperature and $S$ the entropy. Gibbons and Hawking \citep{Gibbons:1977mu} argued that the impossibility of defining energy in dS implies that we are computing a microcanonical partition function, as we are  effectively  fixing the energy to be zero \citep{Galante:2023uyf}.}
\begin{equation}
    e^{S_{\textnormal{dS}}} = \sum_\mathcal{M} \int [ \mathcal{D} g ] e^{-S_E[\Lambda, g_{ij}; \mathcal{M}]} Z_{\textnormal{matter}}[g_{ij};\mathcal{M}] \, ,
    \label{intro:GH}
\end{equation}
where $S_E[\Lambda,g_{ij};\mathcal{M}]$ is the Euclidean gravitational action in a $(d+1)-$dimensional spacetime with $d\geq 2$, positive cosmological constant $\Lambda > 0$ and the sum is over compact manifolds $\mathcal{M}$. This path integral is schematic at best as gravity is a non-renormalisable theory. In fact, the path integral measure can't be properly defined and moreover there is also the problem of the metric conformal mode being unbounded \citep{Gibbons:1978ac}.\footnote{In $2d$ the Euclidean gravitational path integral can be formulated more precisely \citep{Polchinski:1989fn,Distler:1988jt,David:1988hj,Bautista:2019jau}. In the context of dS$_2$  this has been studied in \citep{ Anninos:2021eit,Muhlmann:2022duj,Muhlmann:2021clm}} Despite this, the dominant saddle of (\ref{intro:GH}) is given by the $(d+1)-$sphere, \emph{i.e.} Euclidean $(d+1)-$de Sitter space. Notice that  we also expect the path integral to be   summing over geometries with different topologies. Upon evaluating the on-shell action on the leading saddle we obtain the expected area law 
\begin{equation}
    S_{\text{dS}}^{(0)} = \frac{A}{4 G_{N}}\, .
\end{equation}
here $G_N$ is the Newton constant in $(d+1)$-dimensions. In general, we expect quantum corrections to the entropy arising from both the geometry and the matter content of the theory. In particular, if we restrict  to the sphere saddle, corrections to $S_{\text{dS}}^{(0)}$ arising from quantum fluctuations give rise to logarithmic divergencies which provide unambiguous data that should be reproduced by any microscopic spacetime proposed model  \citep{Anninos:2020hfj}. However, we should remark that at the formal level, this is a delicate issue. UV-divergences arising in the regularisation of these corrections should be absorbed into renormalised coupling constants and common lore tells us that physically meaningful results must be defined in terms of low-energy physical observables invariant under diffeomorphisms and local field redefinitions.  While in flat space we can always rely on the S-matrix to do so  and in AdS we can use the boundary CFT,   for dS  we only have   $S_{\text{dS}}^{(0)}$ as a gauge and field redefinition invariant quantity. Additional observables have been elusive in the de Sitter context and therefore it is not clear how to define sensible coupling constants (see a discussion in appendix I of \citep{Anninos:2020hfj}). 
Finally, as we review below, even if we manage to solve this issue, the Euclidean computation  showcases the appearance of  \emph{edge modes} contributions (new degrees of freedom?) beyond the expected bulk one,  that do not have a global Lorentzian analogue. These results show interesting open questions and interplay between the Group Theory and the Euclidean approaches.  
\\ 

\nin \textbf{Outline \& Results}
\vspace{2mm}

\nin In this note we start an in depth study of the simplest spinning Discrete Series representation of SO(1,4). We first focus on its group theoretical properties taking as a working example the case of the spin$-1$ gauge field. We then path integrate the theory paying special attention to the gauge symmetry and the zero modes, this allow us to compute the $1-$loop partition function and recast the result in terms of Harish-Chandra characters of Discrete Series UIRs of SO(1,4). In addition to the character associated to the bulk degrees of freedom we further identify the \emph{edge mode} character of \citep{Anninos:2020hfj} with that of two $2-$dimensional Discrete Series representations with $\Delta = 1$. We then study the realisation of the UIR in the single particle Hilbert space of a Lorentzian gauge field. To this end we construct the corresponding mode functions and show how the symmetry algebra acts on them. Finally we  show that a proper global gauge invariant treatment of a spin-1 field theory does not account for the edge degrees of freedom. 

\vspace{2mm}

\nin The paper is organised as following, in section \ref{sec:geometry} we review the main geometric properties of de Sitter spacetimes focusing on the group theoretical structure and give a survey of the relevant UIRs appearing in SO(1,4), the general $SO(1,d+1)$ case is left for the Appendix \ref{app:group}. In section \ref{gaugefield} we proceed to study the gauge field in a fixed dS background. We first study the Euclidean picture were computations are clear-cut. Afterwords, we try to reproduce these features from a Lorentzian perspective  by canonically quantising the theory. We find two Discrete Series modules in the single particle Hilbert space of the theory. Details on how the Group Theory constraints the mode solution and the specific identification of all the features is left for the Appendix \ref{S3irreps}. We end up with some discussion, open questions and possible resolutions to the features discussed in the main body of the paper.

\section{Geometry  of dS$_{d+1}$}
\label{sec:geometry}

In this section we present the basic geometric ingredients needed for discussing de Sitter spacetime making special emphasis in group theoretic aspects. 
We consider here de Sitter in arbitrary dimension, whereas in the following sections we focus on dS$_4$. Our presentation is kept minimal when possible, further details can be found in \citep{Anninos:2012qw,Spradlin:2001pw,Sun:2021thf,Dobrev:1977qv,Galante:2023uyf} and appendices \ref{app:group} and \ref{kills}.

The starting point is to view the $(d+1)$-dimensional  de Sitter spacetime (dS$_{d+1}$) as a codimension-1 surface $\Sigma$ embedded in a $(d+2)$-dimensional Minkowski spacetime \citep{Scho}
\begin{equation}
   -X_0^2 + X_1^2 + \cdots + X_{d+1}^2 = \ell^2~~ \hookrightarrow ~~\textnormal{d}s^2 = - \textnormal{d}X_0^2 + \textnormal{d}X_1^2 + \cdots + \textnormal{d}X_{d+1}^2 \, , 
\label{def:emb_mink}
\end{equation}
here $\ell$ is the de Sitter radius that we set equal to $\ell=1$ from now on. The  metric of dS$_{d+1}$ is that induced on $\Sigma$ from  flat   ambient space by solving the constraint in  \eqref{def:emb_mink}. 

From this presentation it is straightforward to recognise the SO$(1,d+1)$ isometry group of dS$_{d+1}$.  As well known, it coincides with the conformal group of $d$-dimensional Euclidean space $\mathbb{R}^d$. A traditional basis for the $\mathfrak{so}(1,d+1)$ Lie algebra is  in terms of antisymmetric generators $J_{AB}=-J_{BA}$ $(A,B=0,1,\cdots, d+1)$ with commutation relations
\begin{equation}
[J_{AB}, J_{CD}] = \eta_{BC} J_{AD} - \eta_{AC} J_{BD} + \eta_{AD} J_{BC}-\eta_{BD} J_{AC} \, , 
\label{def:ds_algebra}
\end{equation}
here $\eta_{AB}=\textnormal{diag}(-+++..)$ is the $(d+2)$-dimensional (ambient) Minkowski metric (\ref{def:emb_mink}). The  quadratic Casimir is defined as
\begin{equation}
    \mathcal{C}_2 \equiv \frac{1}{2} J_{AB} J^{AB} \, .
    \label{Casimir_sec1}
\end{equation}
The generators $J_{AB}$, when realised on fields, can be decomposed as
\be\
J_{AB}=L_{AB}+S_{AB}
\label{Jdec}
\ee
where  
\begin{equation}
     {L}_{AB} =   \left( X_A \partial_{B} - X_B \partial_{A} \right) \, ,
\label{def:KV}
\end{equation}
with $\partial_A=\tfrac\partial{\partial X^A}$ is called the orbital part  and $S_{AB}$, which acts on the field indices,  is called the spinorial part \citep{Dirac:1935zz}. Notice that $L_{AB}$ preserves  the hypersurface \eqref{def:emb_mink}.  

A unitary representation of SO$(1,d+1)$ requires the Lie algebra generators to be realised as anti-hermitian operators on some Hilbert space
\begin{equation}
J_{AB}^\dagger = - J_{AB}\, ,\qquad \forall~A,B \, .
\end{equation} 
It is important to stress that these reality conditions are the ones pertinent for unitary QFTs on a fixed dS$_{d+1}$ background and they differ from the reality conditions traditionally imposed when studying $d$-dimensional unitary conformal field theories in Lorentzian signature \citep{DiFrancesco:1997nk,Dobrev:1977qv,Sun:2021thf}.\footnote{The reality conditions for Lorentzian CFTs are usually worked out in Euclidean signature, they are $$P^\dagger=-K$$
Whereas in the context of the present paper our reality conditions are
$$P^\dagger=-P,\qquad K^\dagger=-K$$ }
The relation between $\mathfrak{so}(1,d+1)$ and the $d$-dimensional Euclidean conformal algebra  plus further details  can be found in  appendix \ref{dsCFT}. 


\subsubsection*{Global Coordinates} 

This is a coordinate chart that covers the full spacetime. It is found by solving the constraint in \eqref{def:emb_mink} as
\begin{equation}
    X^A =(X^0,X^i)= (\sinh T, \cosh T\, n^i ) \, , \quad i = 1, \cdots, d+1\, ,
    \label{res:embed_global}
\end{equation}
with $n^i$ a unit vector in $\mathbb R^{d+1}$. The line element induced from this parametrization is
\begin{equation}
   \text{\sf de Sitter}_{d+1}:~~~ \textnormal{d}s^2 = -\textnormal{d}T^2 + \cosh^2 T\, \textnormal{d}\Omega_d^2 \, ,
   \label{GlobalC}
\end{equation}
here $T \in \mathbb R$ and d$\Omega_d^2$ is the round metric on $S^d$. In this coordinate system, constant time slices are compact spheres that shrink for $T<0$ and  grow for $T>0$. The  $T > 0$ part of the geometry can be seen as   a closed exponentially expanding universe. The expansion rate is set by the de Sitter radius  which was set to $\ell=1$ in \eqref{GlobalC}.  The explicit form of the ten Killing vectors of dS$_4$ can be worked out from (\ref{def:KV}) and (\ref{res:embed_global}) and are spelled out in appendix \ref{kills}. A characteristic feature of de Sitter spacetime  is the absence of a globally defined timelike Killing vector.

\subsubsection*{Conformal Coordinates} 

We can conformally compactify dS making
\begin{equation}
    \cosh T = \frac{1}{\sin t} \, .
\end{equation}
The  metric then takes  the form
\begin{equation}
    \textnormal{d}s^2 = \frac{-\textnormal{d}t^2 + \textnormal{d}\Omega_d^2}{\sin^2 t} \, , \quad t \in (-\pi,0) \, .
    \label{def:ds_conformalmetric}
\end{equation}
In this way, dS is mapped  to a finite strip of Einstein Static Universe (ESU) \citep{Hawking:1973uf}. The late time geometry is located in the $t\to0$ region while the early time geometry corresponds to $t\to-\pi$. This is the coordinate system we will work with.

\subsubsection*{Static patch} 

This coordinate system  describes the region of spacetime causally accessible to a free falling observer. It is given by \citep{Spradlin:2001pw}
\begin{equation}
 \textnormal{d}s^2 = - \cos^2 \rho\, \textnormal{d} \tau^2 + \textnormal{d}\rho^2 + \sin^2 \rho\, \textnormal{d}\Omega_{d-1}^2 \, ,
\label{StatPat}
\end{equation}
here $\tau \in \mathbb{R}$ and $\rho \in [0,\pi/2)$. The observer sits at  $\rho = 0$   and the surface $\rho = \frac{\pi}{2}$ corresponds to the cosmological event horizon surrounding they. Standard horizon arguments\footnote{Such as the absence of   singularities of the Euclidean metric.} associate a temperature to the dS spacetime. Reinserting $\ell$, the de Sitter temperature is  \citep{Gibbons:1977mu}
$$T_{dS}=\frac1{2\pi\ell} \, .$$
The amusing feature of \eqref{StatPat}, in contrast to (\ref{GlobalC}), is the appearance of the time-like killing vector $\bm \partial_\tau$. From the group theory point of view, motions along it  correspond to a $\mathfrak{so}(1,1)\subset \mathfrak{so}(1,d+1)$  non-compact  generator. Yet, this Killing vector becomes spacelike if we extend it beyond the cosmological horizon. This fact precludes a global notion of energy for de Sitter spacetime associated to $\bm\partial_\tau$.

\subsubsection*{Euclidean de Sitter: the Sphere} 

Euclidean de Sitter spacetime can be defined through a Wick rotation of the embedding time coordinate $X^0$ in \eqref{def:emb_mink}. In the global patch \eqref{res:embed_global}, this is achieved writing  $T = i \chi - i \frac{\pi}{2}$, then
\begin{equation}
    \textnormal{d}s_E^2 = \textnormal{d}\chi^2 + \sin^2 \chi\, \textnormal{d}\Omega_d^2 \, , ~\qquad~\chi\in(0,\pi) \, .
    \label{def:ds_eucli}
\end{equation}
Similarly,  for the static patch metric,  making $\tau=i\vartheta$ in \eqref{StatPat} one obtains
\begin{equation}
\textnormal{d}s_E^2 =  \cos^2 \rho\, \textnormal{d} \vartheta^2 + \textnormal{d}\rho^2 + \sin^2 \rho\, \textnormal{d}\Omega_{d-1}^2 \, .
\end{equation}
In both cases, after Wick rotation, the obtained Euclidean de Sitter manifold becomes closed with an induced round $S^{d+1}$ metric. 

Path integral methods  associate an entropy to de Sitter spacetime \citep{Gibbons:1977mu}.  Since the leading saddle point for the gravitational path integral is given by the round sphere $S^{d+1}$, the entropy for  $(d+1)$-dimensional de Sitter spacetime, is found to be\footnote{The absence of any notion of energy in de Sitter spacetime is crucial to the  derivation \eqref{dSentr}.} 
\be
S_{\text{dS}}^{(0)}\Big\rfloor_{_{S^{d+1}}}=\frac A {4G_N} \, , 
\label{dSentr}
\ee
where $A=\text{vol}(\Omega_{d-1})\,\ell^{d-1}$ is the area of the cosmological horizon located at $\rho=\pi/2$ in  \eqref{StatPat}. 
Computing corrections to the leading entropy of de Sitter spacetime is in general not straightforward since we do not know how to properly treat the path integral measure when considering fluctuating geometries. On the other hand, we could study the contribution to the entropy stemming from the fluctuations of quantum fields in a fixed $S^{d+1}$ background. This procedure amounts to compute the $1$-loop partition function of matter fields placed on a round sphere and will be reviewed below (see \citep{Anninos:2020hfj} for a general discussion).

\subsection{Representation Theory}
\label{subsec:reptheory}

The representation theory of the SO$(1,d+1)$ group is a rich and deep subject and there are many classical and recent reviews that describe it in full details \citep{Dobrev:1977qv,Sun:2021thf,ottoson,schwarz}. Stemming from Wigner's classification in flat space, we expect  quantum fields propagating in dS$_{d+1}$ to furnish  UIR's of SO$(1,d+1)$. Here we will follow a minimal approach relevant for the discussion of the main body of the text, leaving some general comments to the appendix \ref{app:group}. 

We follow the traditional notation stemming from the isomorphism between SO$(1,d+1)$ and the $d$-dimensional Euclidean conformal group. The irreducible representations are labelled by weights $(\Delta,\bm s)$ for the $\mathfrak {so}(1,1)\times \mathfrak {so}(d)\subset\mathfrak {so}(1,d+1)$ subgalgebras. They are customarily called conformal weight and  highest weight vector respectively (see app. \ref{app:group} for definitions). It is only for specific values of $(\Delta,\bm s)$ that one obtains  unitary irreducible representations, UIR's for short (see  App. \ref{app:group} for a review of them).

A set of important paradigmatic bosonic examples  which repeatedly show up in discussions of UIRs in dS$_{4}$ are (see the table below)\footnote{General $d$-dimensional considerations are left to the App. \ref{app:group}}
\begin{itemize}
    \item The \emph{Principal Series} $\pi_\nu$: $\Delta = \frac{3}{2} + i \nu$ with $\nu \in \mathbb{R}$. 
    \item The \emph{Complementary Series} $\gamma_\Delta$: $\frac{3}{2} < \Delta < 3$. 
    \item The \emph{Discrete Series} $\mathcal{U}_{s,t}^{\pm}$: 
    ${\bm s }=s$, $t = 0,1,\cdots,s-1$ and $\Delta = 2+t$. 
\end{itemize}
The first two cases displayed here correspond to heavy and light scalar fields respectively, while the third one corresponds to traceless totally symmetric spin-$s$ partially massless gauge field with $2(s-t)$ propagating degrees of freedom \citep{Boers:2013pba},\citep{Hinterbichler:2016fgl}.

\begin{table}[h!]
\begin{center}
\begin{tabular}{ |c|c|c|c|c| } 
\hline
Representation & $\Delta$ & $\bm{s}$ & Field Realisation \\
\hline 
Principal Series: $\pi_\nu$& $\frac{3}{2} + i \nu$, $\nu \in \mathbb{R}$ & 0 & Heavy Scalar field $m \ell > \frac{3}{2}$ \\
Complementary Series: $\gamma_\Delta$ & $\frac{3}{2} < \Delta < 3$ & 0 & Light scalar field $0< m \ell \leq \frac{3}{2}$\\ 
Discrete Series: $\mathcal{U}_{s,t}^\pm$ & $2+t$, $0\le t< s$ & $s$ & (Partially) Massless Gauge Field \\
\hline
\end{tabular}
\caption{Simplest SO(1,4) bosonic UIRs  and their  $\mathfrak {so}(1,1)\times \mathfrak {so}(3)\subset\mathfrak {so}(1,4)$ labels. }
\label{table1}
\end{center}
\end{table}

Some comments are in order, the above UIR's and their extensions to arbitrary $\bm s$ (arbitrary Young tableuxs), including the possibility of fermionic degrees of freedom (see \citep{Basile:2016aen},\citep{Letsios:2023qzq}) have a positive, semi-definite inner product on the space of states (see   Appendix \ref{app:group}). Notice that the Principal Series representation although labelled by a complex scaling dimension $\Delta$, corresponds to a Unitary representation. It is realised in terms of  heavy free scalar fields in a fixed de Sitter background. Thus, we stress that from a group theory perspective,  a complex scaling dimension in the de Sitter context does not imply  a non-unitary theory as it would have been the case if we had a bonafide (Lorentzian) CFT.

The spinning Discrete Series representation in dS$_4$ are  characterised by their spin-$s$  and  an integer $0\le t<s$ customary called \emph{depth}. Fields of maximal depth $t=s-1$ have two propagating degrees of freedom, they are called in the literature \emph{massless gauge fields}. While a massive (\emph{i.e} non gauge invariant) higher spin field will have in general $2s+1$ propagating degrees of freedom, for specific tuning of the mass parameter and the cosmological constant the field equation develops a gauge invariance that reduces the number of propagating degrees of freedom to $2(s-t)$ effectively, these are called  \emph{partially massless fields}. These features of the wave equations were originally discovered for the spin$-2$ particle  in \citep{Deser:1983tm,Deser:1983mm} and later generalised for higher spin fields  in \citep{Higuchi:1986wu,Higuchi:1986py}. Furthermore, as originally recognised by Higuchi, any small deviation from these (discrete) critical masses renders the theory non-unitary through the appearance of ghost-like kinetic terms. These properties make the representation theory for de Sitter more involved than those of flatspace or AdS. 

Furnishing the Discrete Series representations as free quantum fields in a Lorentzian picture is far from trivial. Even for $d=1$ a realisation of the discrete series\footnote{For $d=1$ the discrete series are labelled by $\mathcal{D}^\pm_\Delta$ with $\Delta\in\mathbb N$.} devoid of problems was a subtle issue recently solved in \citep{Anninos:2023lin} (see also footnote 6 in \citep{Anninos:2021ene}). While less generic than the Principal Series, Discrete Series in dS$_2$ generically appear in the multi particle Hilbert space of heavy particles \citep{Dobrev:1977qv}.

The main interest  of the present  note is the field theoretic realisation of the simplest spinning discrete series of SO(1,4), namely $\mathcal{U}_{1,0}^{\pm}$. On general grounds we expect it to be represented by a spin$-1$ gauge field in $3+1$ dimensions (see \citep{Higuchi:1985ad} for previous work). As discussed above this UIR is characterised by the $\mathfrak{so}(1,1)\times\mathfrak{so}(3)$ labels  $(\Delta,\bm s) =( 2,1)$. Since the Discrete Series UIRs is infinite dimensional, its restriction to   SO$(4)$ implies an infinite tower of representations. Indeed, the  SO$(4)$ content of $\mathcal{U}_{1,0}^{\pm}$   read off from the induced representation construction of the UIR results \citep{Dobrev:1977qv,Sun:2021thf} 
\begin{equation}
    \mathcal{U}_{1,0}^\pm \Big \rfloor_{\textnormal{SO}(4)} = \bigoplus_{k \geq 1} \mathbb{Y}_{k,\pm 1} \, .
    \label{eq:SO4content}
\end{equation}
Here $\mathbb{Y}_{k,\pm1}$ denotes the SO$(4)$ irreducible representation labelled by the highest weight vector ${\bm s} = (k,\pm 1),~k\in\mathbb N$. The $\pm 1$ distinguishes the chirality of the representation.\footnote{Chirality relates to the sign of the second Casimir of SO$(4)$ as discussed in detail in Appendix \ref{S3irreps}} We remind the reader that chirality is only present for even dimensional SO$(2r)$ groups, and manifests  in the last component of the highest weight vectors being either positive of negative (see App. \ref{app:group} for more details). From a physical point of view the full orthogonal group O$(4)$ also includes space reflections. These reflections map any given representation of SO$(4)$ to its image $(k,h) \to (k,-h)$, thus, 
any  field theoretic realisation of the Discrete Series should contain both chiral sectors.  

\vspace{2mm}

\nin In the next section we discuss different ways to realise this representation in a field theoretic set up. We first discuss the Euclidean formulation of the theory were we compute the sphere path integral taking into account the contribution of the ghosts and their corresponding zero modes. We review how a careful treatment of the Euclidean theory showcases features of Lorentzian features. We then turn our attention to the Lorentzian realisation putting special emphasis in the Group theoretical constraints and displaying some tensions with the Euclidean perspective.

\section{Linearised Yang+Mills Theory}
\label{gaugefield}

We will start this section focusing on the Euclidean picture where computations are clear-cut. We will discuss the sphere path integral quantisation and perform a one-loop computation of the theory. We will show how Lorentzian physics can be uncovered in an Euclidean computation through a careful treatment of the gauge symmetry and ghost zero modes. While the result of path integration is of course UV-divergent, in $4-$dimensions a universal (logarithmic) coefficient can be extracted,  exactly matching the well known conformal anomaly (see \citep{birrell_davies_1982}). To obtain this result it is crucial to account for the contribution of \emph{edge modes}, which are interpreted as degrees of freedom living in a codimension-$2$ surface. We then turn our attention to the canonical Lorentzian realisation where we put special emphasis in the Group theoretical constraints and show some tensions with the Euclidean picture.

We will consider a massless spin$-1$ Yang-Mills field   on an Euclidean dS spacetime. The Lie algebra   anti-hermitian generators $T^a$ satisfy 
\begin{equation}
[T^a, T^b] = f^{abc} T^c \, .
\end{equation}
The action of this theory is given by
\begin{equation}
S[A] = \frac{1}{2 \textnormal{g}^2} \int_{\mathcal{M}} d^{d+1}x \sqrt{-g} \, \textnormal{Tr} F^2 = \frac{1}{4 \textnormal{g}^2} \int_{\mathcal{M}} d^{d+1}x \sqrt{-g} \, F_{\mu \nu}^a F^{a\, \mu \nu} \, ,
\label{def:actionA}
\end{equation}
where
\begin{equation}
F_{\mu \nu} = \partial_\mu A_\nu - \partial_{\nu} A_\mu + [A_\mu, A_\nu]\, , \qquad A_\mu = T^a A_\mu^a \, ,
\end{equation}
and $\textnormal{g}$ is the YM coupling of the theory. The action (\ref{def:actionA}) is invariant under the gauge transformations 
\begin{equation}
A_\mu \to A_\mu + \partial_\mu \alpha + [A_\mu, \alpha] \, ,
\label{def:gaugetrans}
\end{equation}
where $\alpha = \alpha^a T^a$. Quantising this theory is a standard exercise, in the following  we will review different approaches showcasing relevant features that show up for a fixed de Sitter background. At $1$-loop order the gauge field self-interactions  can be neglected, thus, the abelian and non-abelian cases become equivalent. They  differ only by the colour degree of freedom that  amounts simply to a $N=\textnormal{dim}(G)$ factor. Effectively we will be computing a path integral over a U$(1)$ gauge field.

\subsection{Path Integral Approach}
\label{secPI}

As emphasised above, Wick rotating dS to Euclidean signature results in the round sphere metric 
(\ref{def:ds_eucli}). Our main interest is then the path integral quantisation and the $1$-loop partition function of a Yang+Mills theory $A^a_\mu$ with gauge group $G$ placed on a round sphere
\begin{equation}
\mathcal{Z} = \int \frac{\mathcal{D}A^a}{\textnormal{Vol}(G)_{\text{local}}} e^{-S_E[A^a]}\, .
\label{PI_first}
\end{equation}
The Vol$(G)_{\text{local}}$ factor  is included by hand to quotient the overcounting of field configurations when performing the $A^a$-path integral. As we will discuss below the inclusion of this factor is also fundamental to maintain the locality of the theory.

A proper path integral quantisation of the theory requires gauge fixing to avoid overcounting gauge equivalent configurations. The standard way to proceed is either by introducing Fadeev-Popov ghosts or by a BRST treatment of the gauge symmetry. Here we will take proper account of gauge invariance following the geometric approach developed in \citep{BABELON1979246,MAZUR1990187,bern}. We start parametrising the field as
\begin{equation}
A_\mu = A_\mu^T + \partial_\mu \chi \, ,
\label{eq:aparamet}
\end{equation}
with $A_\mu^T$ the transverse component of $A_\mu$ satisfying $\nabla^\mu A_\mu^T = 0$, and $\partial_\mu\chi$  the  longitudinal (pure gauge) component of the field. An important observation to be discussed below is that  different constant values $\chi(x)=\chi_0 $  (zero modes of the scalar Laplacian) give the same gauge field configuration $A_\mu$, hence integration over the zero mode $\chi_0$ should be excluded from the $\chi$-path integral. We will denote with a prime the fact that we are not integrating over these modes.

We exploit the fact that $A_\mu^T$ is transverse by introducing the following operator
\begin{equation}
    D_{\mu \nu} = -\nabla^2_{(1)} \delta_{\mu \nu} + R_{\mu \nu} \, ,
    \label{commutadorA}
\end{equation}
Here $\nabla^2_{(1)} = \nabla^\mu \nabla_\mu$ is the Sphere Laplacian   acting on vectors and $R_{\mu \nu}$ is the Ricci curvature tensor satisfying  
\begin{equation}
    [\nabla_\mu, \nabla_\nu] A^\nu = -R_{\mu \nu} A^\nu \, .
    \label{richo}
\end{equation}
All in all the Yang+Mills action reduces to
\begin{equation}
    S_E[A^T] = \frac{1}{2\textnormal{g}^2} \int_{S^{d+1}} d^{d+1}x \sqrt{g}\,  A_\mu^T\left( -\nabla^2_{(1)} + d \right)A^{T \mu} \, ,
    \label{transverseAction}
\end{equation}
where we have used that $A_\mu^T$ is transverse, the operator (\ref{commutadorA}), and the fact that $S^{d+1}$ is a maximally symmetric space. The path integral \eqref{PI_first} is now given by 
\begin{equation}
    \mathcal{Z} =  \int \frac{\mathcal{D}A^T \mathcal{D}' \chi}{\textnormal{Vol}(G)_{\text{local}}} \,  J \, e^{-S_E[A^T]} \, , 
\end{equation}
where $J$ is the Jacobian arising from the change of variables $A\mapsto(A^T,\chi)$  in (\ref{eq:aparamet}). Notice that the pure gauge modes decouple from the action (\ref{transverseAction}) yet they are relevant to fix the value of $J$. To do so we demand the parametrisation (\ref{eq:aparamet}) to be consistent with the normalisation condition
\begin{equation}
    1 = \int \mathcal{D}A e^{-\frac{1}{2\textnormal{g}^2}(A,A)} \, .
    \label{normA}
\end{equation}
Performing the change of variables \eqref{eq:aparamet} in this expression   we obtain
\begin{equation}
    J = \det{'} \left( -\nabla^2_{(0)} \right)^{1/2} \, ,
    \label{jacobian_chi}
\end{equation}
Introducing two fermionic fields $b, \bar{b}$, we can express the determinant  (\ref{jacobian_chi}) as the path integrated result of the following ghost action
\begin{equation}
    S_{gh}[\bar{b},b] =  \int_{S^{d+1}} d^{d+1}x \sqrt{g}\,\bar{b}\nabla^{'2}_{(0)} b \, .
\end{equation}
As a result the path integral reads
\begin{equation}
    \mathcal{Z} =  \int \frac{\mathcal{D}\bar{b} \mathcal{D}b \mathcal{D}'\chi \mathcal{D}A^T}{\textnormal{Vol}(G)_{\text{local}}} e^{-\left( S_E[A^T] + S_{gh}[\bar{b},b] \right)} \, .
    \label{PIfinal}
\end{equation}
A few comments are in order, the factor Vol$(G)_{\text{local}}$ is theory dependent and is formally given by the volume of the (local) gauge transformations. In general it is given by an integral over $n=\textnormal{dim}( G)$ local scalar fields. The   parametrisation (\ref{eq:aparamet}) for the gauge field $A_\mu$ decouples the gauge degrees of freedom $\chi$,   thus we expect the measure of $\chi$ to exactly cancel the factor Vol$(G)_{\text{local}}$. However, as discussed previously, we have substracted the constant zero-mode $\chi_0$ from the measure $\mathcal{D}' \chi$. As a result, the cancellation is not complete an a mismatch arises since  constant modes are present in the full group volume. In fact, the constant modes are the ones that generate the global symmetries of the theory. Notice that this in sharp contrast to the case of AdS of flat space in which standard boundary conditions for the field configurations forbid such global features to survive.\footnote{In AdS this term also survives if we consider Neumann boundary conditions, see \citep{Anninos:2020hfj}.} We thus expect to have a residual (global) group  volume associated to the global symmetries given roughly by 
\begin{equation}
Z_G=\frac{\int \Pi_{a=1}^{\textnormal{dim}G}\mathcal{D}'\chi^a}{\textnormal{Vol}(G)_{\text{local}}}\sim \frac{1}{\textnormal{Vol}(G)_{\textnormal{Global}}}  \, . 
\end{equation}
We will make a few comment on this result below. A detailed discussion can be found in \citep{Anninos:2020hfj}. 
 
To compute the path integral \eqref{PIfinal}, 
the $A_\mu^T$ field configurations (transverse spin-$1$) can be decomposed in terms of transverse vector spherical harmonics (see \citep{Higuchi:1986wu} and App.\ref{S3irreps} for the case of the 3-sphere), while the longitudinal (pure gauge) modes and the   local scalar fields generating the Vol$(G)_{\text{local}}$ factor can be expressed in terms of scalar harmonics and derivatives thereof. As a result,
the sphere $1-$loop partition function becomes
\begin{equation}
\mathcal{Z} = Z_G \left( \frac{\det'(-\nabla^2_{(0)})^{1/2}}{\det\left( -\nabla^2_{(1)}+d \right)^{1/2}} \right)^{\textnormal{dim}(G)} \, .
\label{PI_completa}
\end{equation}\\

\nin In the following we proceed now to compute the determinants following standard Heat Kernel techniques \citep{vassilevich}.  We refer the reader to \citep{Anninos:2020hfj} for the computation of the local gauge group volume $Z_G$.\footnote{While normally overlooked, the volume of the gauge group $\textnormal{Vol}(G)_{\text{local}}$ plays a crucial role for consistency with locality and unitarity of the theory.} We start rewriting the path integral \eqref{PI_completa} as
$${\cal Z}=Z_G \,\left({\cal Z}_{(1)}{\cal Z}_{(0)}\right)^{{\textnormal{dim}(G)}}$$
where 
\begin{equation}
\begin{split}
    \log \mathcal{Z}_{(1)} &= \int_0^{\infty} \frac{d\tau}{2\tau} \, e^{-\epsilon^2/4\tau} \,  \textnormal{Tr} e^{-\tau\left( -\nabla^2_{(1)}+d \right)} \\ 
    \log \mathcal{Z}_{(0)} &= -\int_0^{\infty} \frac{d\tau}{2\tau} \, e^{-\epsilon^2/4\tau} \,  \textnormal{Tr}' e^{-\tau\left( -\nabla^2_{(0)} \right)} \, .
    \label{res:gauge_ghost_pf}
\end{split}
\end{equation}
Here, the prime $'$ in ${\cal Z}_{(0)}$ denotes the absence of the zero mode contribution when computing the trace. As usual, the insertion $\exp(-\epsilon^2/4\tau)$  regulates the UV-divergence when $\tau\to0$. 

To compute the traces in \eqref{res:gauge_ghost_pf} we need the details of the complete basis of spin$-1$ and spin$-0$ symmetric traceless spherical harmonic as well as the corresponding degeneracies for each eigenvalue. The general problem of constructing such basis for arbitrary spin$-s$ tensor fields on spheres was carried in \citep{Higuchi:1986py} and we outline the basic ingredients in the appendix \ref{app:stsh_modes}. For the $s=0$ case we have
\begin{equation}
    -\nabla^2_{(0)} \psi_{n} = \lambda_n \psi_{n} \, , \qquad \lambda_n = n(n+d)\, ,~~n=0,1,2,...
\end{equation}
with degeneracies given by
\begin{equation}
    D_n^{d+2} = {n+d+1 \choose d+1} - {n+d-1 \choose d+1} \, .
    \label{Ds}
\end{equation}
Now the trace can be performed 
\begin{equation}
    \log \mathcal{Z}_{(0)} = - \int_{0}^\infty \frac{d\tau}{2\tau} e^{-\epsilon^2/4\tau} \sum_{n=1}^\infty D_{n}^{d+2} e^{-\tau n (n+d)} \, ,
\end{equation}
notice that the sum starts at $n=1$ since we are instructed to avoid the constant zero mode associated to $n=0$. Completing the square in the exponential and performing a Hubbard-Stratonovich trick via an auxiliary variable $u$ we obtain
\begin{equation}
    \log \mathcal{Z}_{(0)} = -\int_0^\infty \frac{d\tau}{2\tau} e^{-\epsilon^2/4\tau} e^{\tau\frac{ d^2}{4}} \int_{\mathbb{R}+i\delta} du \frac{e^{-u^2/4\tau}}{\sqrt{4\pi \tau}} f(u) \, ,
    \label{trick}
\end{equation}
where
\be
\quad f(u) = \sum_{n=1}^\infty D_n^{d+2} e^{i u(n+\frac{d}{2})}\,.
\label{fu}
\ee
This sum can be performed
$$f(u)=e^{  i\tfrac d2 u}\left[\frac{1+e^{iu}} {(1-e^{iu})^{d+1}}-1\right]$$
The second term in the bracket can be interpreted as the subtraction of the zero mode when computing the trace. Finally, the $u$-integral contour in \eqref{trick} is done slightly above the real axis ($\delta>0$), in order to avoid divergences from the denominators in $f(u)$ . 

The $\tau$ integral can now be obtained by analytic continuation in $d$ to give
\begin{equation}
     \log \mathcal{Z}_{(0)} = -\int_{\mathbb{R}+i\delta} du f(u) \frac{e^{-i \frac{d}{2}\sqrt{u^2+\epsilon^2}}}{2 \sqrt{u^2+\epsilon^2}} \, ,
\end{equation}
where we have assumed that $0<\delta<\epsilon$.
Calling $u = i t$ and deforming the original contour to the branch cut located at $u=i\epsilon$ one finds \citep{Anninos:2020hfj} 
\begin{align}
    \log \mathcal{Z}_{(0)} &= -\int_{\epsilon}^\infty \frac{dt}{2t} f(q) \left( q^{\frac{d}{2}} + q^{-\frac{d}{2}} \right) \, ,  
    \label{ref:ghost_contr}\\
    &= -\int_{\epsilon}^\infty \frac{dt}{2t} \left[ \frac{1+q}{1-q} \frac{q^{\frac{d}{2}}}{(1-q)^d} - q^{\frac{d}{2}}\right] \left( q^{\frac{d}{2}} + q^{-\frac{d}{2}} \right) \, .
    \label{PI_gaugemodes}
\end{align}
where $q = e^{-t}$.  It is important to remember that the second term in the brackets arises from the subtraction of the zero mode contribution.

Performing similar steps for the  spin-1 piece in (\ref{res:gauge_ghost_pf}) with $\lambda_n=n(n+d)-1+d$ one finds
\begin{equation}
    \log \mathcal{Z}_{(1)} = \int_0^\infty \frac{\textnormal{d}\tau}{2\tau} e^{-\epsilon/4\tau} e^{\tau \left(\frac{d}{2}-1\right)^2} \int_{\mathbb{R}+i\delta} \textnormal{d}u \frac{e^{-u^2/4\tau}}{\sqrt{4\pi t}} f_1(u)
\end{equation}
with
$$f_1(u) = \sum_{n=1}^\infty D_{n,1}^{d+2} e^{iu(n+\frac{d}{2})}\, .$$
Notice (\ref{eigenvalues}) implies  there are no zero modes for transverse vectors on the sphere. The transverse spin-1 degeneracies $D_{n,1}^{d+2}$ can be found in \eqref{eigenvalues} and obey the relation
\begin{equation}
    D_{n,1}^{d+2} = D_{n}^{d+2} D_1^d - D_0^{d+2} D_{n+1}^d \, .
    \label{res:branching_deg}
\end{equation}
where $D^{d+2}_n$ can be read in \eqref{Ds}. As a result, a closed expression for $f_1$ can be found
\be
f_1(q)= \frac{1+q}{1-q}\left(d\frac{q^{\frac{d}{2}}}{(1-q)^d}-\frac{q^{\frac{d-2}{2}}}{(1-q)^{d-2}}\right)+q^{\frac{d-2}{2}}
\ee
here we defined $q=e^{iu}$. The last term in the bracket leads to a $1/t$ contribution in the small$-t$ limit and implies that there is a logarithmic UV divergence that can be associated to the cutoff scale given by the heat-kernel regularisation \citep{Anninos:2020hfj}. As discussed in detail in that reference, this term cancels against a contribution coming from the longitudinal path integral, i.e. $Z_G$. Deforming the $u$-contour, writing $u=it$ and analytically extending in the dimension, the final result is
\begin{equation}
    \log \mathcal{Z}_{(1)} = \int_\epsilon^\infty \frac{dt}{2t} \left[ \frac{1+q}{1-q} \left( d \frac{q^{\frac{d}{2}}}{(1-q)^d} -  \frac{q^{\frac{d-2}{2}}}{(1-q)^{d-2}}  \right) + q^{\frac{d-2}{2}} \right] \left( q^{\frac{d-2}{2}}+q^{-\frac{d-2}{2}} \right)
\end{equation}
Combining the contributions from the determinants  we find  
\begin{equation}
    \log \left(\mathcal{Z}_{(1)}\mathcal{Z}_{(0)}\right) = \int_{0}^\infty \frac{dt}{2t} \left( \frac{1+q}{1-q} \left[ \left( d \frac{q^{d-1}+q}{(1-q)^d} - \frac{q^{d}+1 }{(1-q)^d} \right) -  \left( \frac{q^{d-2}+1}{(1-q)^{d-2}} \right) \right] +  ( 2+q^{d}+q^{d-2}) \right)  \,.
    \label{res:SPI}
\end{equation}

\nin In the following we will interpret \eqref{res:SPI} from a Lorentzian perspective. To this end the terms inside the brackets have been split in two pieces: the two terms in the first parentheses will be associated to  $d-$dimensional contributions coming from the transverse modes and the ghost fields, while the second piece, originally arising  from the transverse modes, will be associate to a $(d-2)-$dimensional contribution whose interpretation will be discussed below. 
\vspace{2mm}

\subsection*{Harish-Chandra characters of SO(1,$d+1$)} 

For any unitary representation $R$ of a semisimple Lie group $G$ one can associate a Harish-Chandra character $\chi_R(g)$ that encodes the information of the given representation in a simple function \citep{harish1,harish2,atiyah}.\footnote{For a finite dimensional representation $R$ of $G$ we define the character associated to an element $g \in G$ as the trace of $R(g)$ over the representation space. For infinite dimensional representations the notion of trace is more subtle and the general theory was developed by Harish-Chandra.}  

For   SO$(1,d+1)$ UIRs, the Harish-Chandra characters were computed in \citep{hiraiI,hiraiII,hiraiIII,Basile:2016aen}. That associated to the spin-1 Type II Exceptional Series  for $d > 3$, which in $d = 3$ reduces to the $s =1$ Discrete Series, is
\begin{equation}
    \chi^{\text{II}}_{ s=1}(t) = d \frac{q^{d-1}+q}{(1-q)^d} - \frac{q^{d}+1 }{(1-q)^d} + 1\,,
    \label{bulk_character_PI}
\end{equation}
which, from now on, we  call  the bulk contribution $\chi^{\text{bulk}}(q)= \chi^{\text{II}}_{ s=1}(t)$. It is straightforward to see it correspond to the first parentheses in (\ref{res:SPI}). It is illuminating to define an \emph{edge} character as 
\begin{equation}
    \chi^{\textnormal{edge}}(t) =  \frac{q^{d-2}+1}{(1-q)^{d-2}} -1 \, .
    \label{edge_character_PI}
\end{equation}
Then, it is remarkable to notice that if we restrict to $d =3$, the \emph{edge} contribution reduces to 
\begin{equation}
    \chi^{\textnormal{edge}}(t) \rfloor_{_{d=3}} = 2 \frac{q}{1-q}
\end{equation}
This expression coincides with the Harish-Chandra character of a SO(1,2) representation $R$ composed of two Discrete Series ${\cal D}^\pm_1$ 
(cf. \citep{Anninos:2023lin})
$$\chi^{\textnormal{edge}}(t) \rfloor_{_{d=3}} =  \chi_{R} (t)\,,~~~~~R_{\text{SO(1,2)}}=\mathcal{D}^+_{\Delta}\oplus \mathcal{D}^-_{\Delta}~~\text{with}~~\Delta=1 .$$

The Harish-Chandra characters   (\ref{bulk_character_PI})-(\ref{edge_character_PI}) correspond  to group elements of the form $g = e^{-it H}$ with $H$ a hermitian $\mathfrak{so}(1,1)\subset\mathfrak{so}(1,d+1)$ generator. It is important to recognise that $\bm H=i\bm D$ becomes  associated to the timelike Killing vector $i\bm\partial_\tau$ of the static patch of de Sitter spacetime (cf. \eqref{StatPat}). Thus, we conclude that the Euclidean path integral on $S^{d+1}$ encodes thermal information even without the presence of a thermal cycle. 

We can now recast the integrand in (\ref{res:SPI}) as
\begin{equation}
    F_{\chi} (q) = \frac{1+q}{1-q} \left( \chi ^{\textnormal{bulk}}(q) - \chi^{\textnormal{edge}}(q) - 2 \right) \, .
\end{equation}
While this explicit form in terms of Harish-Chandra characters was discussed just for the spin-1 field, the result generalises for arbitrary spin-$s$ \citep{Anninos:2020hfj}. 

The last term of (\ref{res:SPI}) is problematic as it has a $q^0$ term which causes the integral to be logarithmically divergent for $t \to \infty$ even in odd spacetimes dimensions where a manifestly covariant local QFT path integral on the sphere cannot have such pathologies. As mentioned before, this contribution stems from zero modes in the original path integral associated to $\chi_0=const.$ As shown in Appendices F \& G in \citep{Anninos:2020hfj},  the log divergencies arising from these modes cancel against an identical contribution coming from Vol$(G)_{\textnormal{local}}$.  The computation of this volume is far from trivial since it has to be computed with the path integral induced metric rather than the usual canonical volume of the group that is defined with respect to the invariant metric (see also \citep{Donnelly:2013tia}). Taking $G = \textnormal{U}(1)$, the gauge group volume contribution gives  \citep{Donnelly:2013tia,Anninos:2020hfj,Giombi:2015haa}
\begin{equation}
    Z_G\, {\exp \left( \int \frac{\textnormal{d}t}{2t}(2+q^d + q^{d-2}) \right)}  = \frac{\textnormal{g}}{\sqrt{2\pi \textnormal{Vol}(S^{d+1})}} \, . 
    \label{prefactor}
\end{equation}
All in all, the $1-$loop sphere partition function of an U$(1)$ gauge field is given by
\begin{equation}
    \mathcal{Z}_{\text{U(1)}, d=3} = \frac{\textnormal{g}}{\sqrt{2\pi \textnormal{Vol}(S^{d+1}})} \exp\left( \int \frac{\textnormal{d}t}{2t} F_{\chi}(q) \right) \, ,
    \label{PI_FINAL}
\end{equation}
with 
\begin{equation}
    F_{\chi}(q) = \frac{1+q}{1-q} \left( \chi ^{\textnormal{bulk}}(q) - \chi^{\textnormal{edge}}(q) - 2 N^{KT}_1 \right) \, , 
\end{equation}
here $N_1^{KT}=1$ corresponds to the number of spin-0 Killing on $S^{d+1}$. 

Let us comment on some features of this result.  Even though we started with an Euclidean path integral on an $S^4$, the final result was shown to contain two different contributions: one (\ref{bulk_character_PI})  associated to the spin-1 Discrete Series representation,  the physical propagating degrees of freedom in the Lorentzian perspective, and the second  (\ref{edge_character_PI}),  associated to Discrete UIRs  of SO$(1,2)$. This last contribution is interpreted as degrees of freedom localised in a co-dimension $2$-surface  (the cosmological horizon in the Lorentzian perspective).

While this sphere path integral and characters are normally UV-divergent it is still possible to extract an universal contribution. Taking a $t \to 0$ expansion will yield terms that scale as $t^{-r}$ with $r>1$ and are associated to local UV divergencies. As it is well known these coefficients can be modified by a local renormalisation of the theory and are thus not meaningful. On the other hand, in even dimensions the $\log$-divergent term is scheme independent and becomes then universal, taking $d=3$ and $t\to0$ in (\ref{res:SPI}) 
\begin{equation}
    \log \mathcal{Z} \simeq \cdots + \log \epsilon \left( -\frac{16}{45} - \frac{1}{3} \right) + \cdots
\end{equation}
where the contributions stem from the bulk and edge character (\ref{bulk_character_PI}) (\ref{edge_character_PI}) respectively. This result exactly matches the expectation from the study of conformal anomaly of a spin$-1$ field in curved backgrounds of \citep{birrell_davies_1982}. Even more, it shows that usual entanglement entropy computations \citep{Casini:2019nmu,Dowker:2010bu} do not account for the contribution of the edge modes. This contribution stemming from a co-dimension $2-$surface appears naturally from a proper treatment of the sphere path integral, and is non-zero for general fields with a gauge symmetry \citep{Anninos:2020hfj}, yet it is still an open problem to realise this contribution from a Lorentzian perspective.
   
\subsection{Hilbert Space/Lorentzian picture}

In this section we discuss the Lorentzian realisation of a $U(1)$ gauge field. Following the previous sections we expect it to carry the simplest Discrete Series representation of SO$(1,4)$, $\mathcal{U}^\pm_{1,0}$, and as such having the SO$(4)$ content displayed in (\ref{eq:SO4content}). A nice set of coordinates that makes manifest the maximally compact subgroup are either the conformal (\ref{def:ds_conformalmetric}) or the global ones (\ref{GlobalC}). Both choices have a $S^3$ that allows a natural action of the SO$(4)$ group. In addition, $4-$dimensional Maxwell theory is known to be classically invariant under Weyl transformations
\begin{equation}
    g'_{\mu\nu}(x)=e^{2\varphi(x)}g_{\mu\nu}(x),~~~A'_\mu(x)=A_\mu(x)
\end{equation}
with  $\varphi(x)$ an  arbitrary function. We further exploit this fact by working in the conformal coordinates (\ref{def:ds_conformalmetric}).   
 
\subsection*{Classical solution}

Choosing the de Sitter metric in conformal coordinates \eqref{def:ds_conformalmetric}, the Maxwell action (\ref{def:actionA})  in dS reduces to that in ESU,
\begin{equation}
S[A] = \int d^{4} x \sqrt{-h} \left(-\frac{1}{4} h^{\mu\alpha}h^{\nu\beta}F_{\mu \nu} F_{\alpha\beta} \right) \, ,
\end{equation}
with $h_{\alpha \beta}$  the metric of Einstein Static Universe (ESU$_4$)
\begin{equation}
ds^2_{ESU}=h_{\mu\nu}\textnormal{d}x^\mu \textnormal{d}x^\nu= -\textnormal{d}t^2 + \textnormal{d}\Omega_3^2 \, .
\end{equation}
Maxwell equations are easily obtained
\begin{equation}
{\sf D}_\mu F^{\mu \nu}=0~~\leadsto~~\partial_\mu(\sqrt{-h}h^{\mu\alpha}h^{\nu\beta}F_{\alpha\beta})=0 \, ,
\label{eomesu}
\end{equation}
here $\sf D_\mu$ denotes the covariant derivative in ESU. We can rewrite Maxwell equation as
\be
\square_{ESU}A_\nu-{\sf D _\nu (D}_\alpha  A^\alpha)-R_{\nu\alpha}[h] A^\alpha=0
\label{box}
\ee
with $\square_{ESU}=h^{\mu\nu}\sf D_\mu D_\nu$ the ESU Laplacian on spin-1 tensors. The ESU Ricci tensor takes the form
\be
R_{\mu\nu}[h]=\textnormal{diag}(0,2\,\mathfrak g_{ab})
\label{Resu}
\ee
with $a,b=1,2,3$ denoting the spatial components and $\mathfrak g_{ab}$ is the 3-sphere metric.\footnote{For completeness we quote that in ESU: ${\sf D}_t=\partial_t$ and ${\sf D}_a=\nabla_a$ where $\nabla_a$ is the covariant derivative on $S^3$.}
 
\vspace{2mm}

\nin We now turn to the gauge fixing issue. We will choose to work in temporal gauge,
\be
\text{\sf Temporal gauge}:~~A_t = 0\,.
\label{Tg}
\ee
This choice does not  fix the gauge completely as we still have the possibility of $t$-independent (residual) gauge transformations 
\begin{equation}
    A_\mu \rightarrow A_\mu + \partial_\mu \alpha(\bm{x}) \, .
    \label{res:residual_gauge}
\end{equation}
here $\bm x$ denotes spatial coordinates. Of course this remaining freedom must be eliminated. To this end, we consider the $A_t$ component from (\ref{box}) in the temporal gauge, 
\begin{equation}
\partial_t\left(\partial_a(\sqrt{\mathfrak g}\mathfrak g^{ab}A_b)\right)=0~~\leadsto~~\partial_t( \nabla_a     A^a )= 0~~\leadsto~~ \bm\nabla \cdot \bm A=f( \Omega)
\end{equation}
with 
$\nabla_a$ its Levi-Civita connection and  $f(  \Omega)$ an arbitrary function on $S^3$. Here we have compactly denoted the spatial components of the gauge field as $\bm A$.  We can use the residual gauge  freedom (\ref{res:residual_gauge}) to impose
\begin{equation}
\text{\sf Coulomb gauge}:~~\bm\nabla \cdot \bm A =  0 \, .
\label{transv}
\end{equation}
In this way we end up with the two expected transverse modes for the gauge field. Note that imposing $A_t=0$ and $\nabla\cdot\bm A=0$ completely fixes the gauge freedom since any gauge transformations preserving \eqref{transv} requires $\tilde\alpha(\Omega)$ to satisfy 
\begin{equation}
    \nabla^2\tilde\alpha(  \Omega)=0 \, .
\end{equation}
On a $3-$sphere the only solution to this equation is   $\tilde\alpha(\Omega)=const.$ As mentioned in the paragraph below eqn. \eqref{eq:aparamet}, this transformation is not present in the theory as it does not move $A_\mu$ along the gauge orbit. 

The equations   for the spatial components of the gauge field are found by  setting $\nu=a$ in \eqref{box}. Notice that our gauge fixing implies that the Lorentz condition is satisfied ${\sf{D}}_\mu A^\mu=0$, then
\begin{equation}
-\partial_t^2 A_a + \nabla^2_{(1)} A_a - 2 A_a = 0 \, ,\quad \nabla_a A^a = 0 \, ,
\label{EsuA}
\end{equation}
here $\nabla^2_{(1)}$ is the 3-sphere Laplacian acting on vectors and we used \eqref{Resu}. To find the mode solution to (\ref{EsuA}) we insert  the ansatz 
\be
\bm A(t, \Omega)=e^{-i\omega t}\bm S( \Omega)
\label{ans}
\ee
and find
\be
(-\nabla_{(1)}^2+2)\bm S=\omega^2\bm S\,.
\ee
This equation says $\bm S$ must be a vector spherical harmonics on $S^3$. Symmetric tensor spherical harmonics on the $n-$sphere are well known and can be found in \citep{Higuchi:1986wu}. For a spin$-1$ tensor three independent solutions appear (see Appendix \ref{eigenD} for details). Among these  only two $i = 1,2$ satisfy the transversality condition
\begin{equation}
    -\nabla_{(1)}^2 { S}_{(i)a} = \lambda {  S}_{(i)a} \, , \qquad \nabla^a {  S}_{(i)a} = 0 \, ,
    \label{S3eigenmal}
\end{equation}
here  $a=1,\cdots 3$ denotes spatial vector components on $S^3$ and
\be
\lambda= (k + 1)^2 - 2\,.
\ee
From a group theory point of view we expect the eigenfunctions $\bm S_{(i)}$ to furnish unitary irreducible representation of  the isometry group of $S^3$, namely  SO$(4)$. These are labelled by the two SO$(4)$ Casimirs that can be written in terms of the anti-Hermitian generators $K_{ij}$ (see Appendix \ref{S3irreps} for further details and definitions) as
\begin{equation}
    \mathcal{C} = \frac{1}{8} K_{ij}K^{ij} \, , \qquad \tilde{\cal C} = \frac{1}{32} \epsilon^{ijkl} K_{ij} K_{kl} \, ,~~~~~i,j,..=1,..,4
\end{equation}
where $\epsilon^{ijkl}$ is the $4-$index Levi-Civita symbol. As discussed in the appendix, they can be related to differential operators in terms of the Killing vector fields of $S^3$. For a spin$-1$ $A^a$ field one finds the relations (\ref{CasiA}), (\ref{Casi2})
\begin{equation}
    \mathcal{C} A^a = \left( \nabla_{(1)}^2 - 2\right) A^a \, , \qquad \tilde{\cal C}A^a =\frac {\epsilon^{abc}}{\sqrt{\mathfrak g}}\partial_b A_c \, ,
    \label{CasimirsMain}
\end{equation}
where we can recognise $\mathcal{C}$ as the equation of motion (\ref{EsuA}) and $\tilde{\mathcal{C}}$ as an helicity operator \citep{Higuchi:1991tn}.\footnote{The appearance of a \emph{mass}-looking term on $\mathcal{C}$ for $s\neq 0$ is generally expected from the square of the spinorial matrix $S_{ij}$ in (\ref{Jdec}).} From this perspective we learn that the traditional vector harmonics $\bm S_{(i)}$ ($i=0,1,2$) appearing in the literature \citep{Higuchi:1986wu,Lindblom:2017maa} are eigenvectors of $\cal C$, but do not diagonalise $\tilde{\mathcal{C}}$. The appropriate $\tilde{\cal C}$ eigenvectors    are   \citep{Higuchi:1991tn} 
\begin{equation}
    {\mathcal{S}}^{klm}_{(\pm)a}( \Omega) \equiv \frac{1}{\sqrt{2}} (  S^{klm}_{(2)a}( \Omega) \pm   S^{klm}_{(1)a }(  \Omega)) \, ,
\end{equation}
where $k,l,m$ denote quantum numbers. Defined  this way they satisfy
\begin{equation}
    \mathcal{C} \bm{\mathcal{S}}^{klm}_{(\pm) }(\Omega) = -(k+1)^2 \bm{\mathcal{S}}^{klm}_{(\pm) }(\Omega)\, , \qquad \tilde{\mathcal{C}} \bm{\mathcal{S}}^{klm}_{(\pm) }(\Omega) = \mp (k+1) \bm{\mathcal{S}}^{klm}_{(\pm) } (\Omega) \, ,
    \label{casis}
\end{equation}
thus $\bm{\mathcal{S}}^{klm}_{(\pm)}$ carry representations of SO$(4)$ with identical $\cal C$ differing by their chirality which is computed by $\tilde{\cal C}$,  see Appendix \ref{S3irreps} for further definitions and derivations. 

The relation between the Casimirs quantum numbers   and the fact that a transverse vector $\bm A$ carries two UIRs can be understood by noticing that \eqref{EsuA} can be written as \citep{Gibbons:1979kq}
\be
(\partial_t+i \nabla\times)(\partial_t-i \nabla\times)\bm A=0,~~~\nabla\cdot\bm A=0
\ee 
where
\be 
(\nabla\times\bm A)^a\equiv \frac{\epsilon^{abc}}{\sqrt{\mathfrak g}}\partial_bA_c\,.
\ee 
Thus, the general solution to \eqref{EsuA} is
\be
\bm A=\bm{\mathcal{A}} _{(+)}+\bm{\mathcal{A}}_{(-)} \, ,
\ee
where 
\be 
\partial_t\bm{\mathcal{ A}}_{(\pm)}=\pm i\, \nabla\times\bm{\mathcal{A}}_{(\pm)} \, . 
\label{EiBjuntitos}
\ee 
This equation can be solved by virtue of $\bm{\mathcal{S}}^{klm}_{(\pm) }$ being eigenvectors of $\bm\nabla\times$ the Separating variables as in \eqref{ans} the solution is
\be
\bm{\mathcal{ A}}_{(\pm)}=\alpha\,e^{-i\omega t}\bm{\mathcal{S}}^{klm}_{(\pm) }(\Omega)+\beta e^{i\omega t}\bm{\mathcal{S}}^{klm}_{(\mp) }(\Omega)
\ee
with $\alpha,\beta$ arbitrary constants. The second equation in \eqref{casis} implies the dispersion relation 
\be
\omega(k) = k +1\,.
\ee
Summarizing, the general solution to the gauge field  equations of motion (\ref{EsuA}) results 
\begin{equation}
    \bm A (t,\Omega) = \sum_{\sigma=\pm} \sum_{klm} \mathcal{N}_k \left( e^{-i \omega t} \bm{\mathcal{S}}_{(\sigma)}^{klm} (\Omega) a_{(\sigma )}^{klm}  + e^{i\omega t}\bm{\mathcal{S}}_{(\sigma) }^{*\, klm} (\Omega) a^{\dagger\,klm}_{(\sigma)}  \right) \, .
    \label{canonical_modes}
\end{equation}
with $\omega(k)= (k +1)$. It is now straightforward to check that constructed like this (\ref{canonical_modes}) also satisfies the Maxwell equations in de Sitter
\be 
\big[D_\mu  F^{\mu\nu}=0~\oplus~A_t=0~\oplus~\nabla\cdot\bm A=0 \big] ~~\leadsto~~(\square_{dS} - 3) A_\mu =0
\label{AdSbox}
\ee 
here $D_\mu$ denotes the de Sitter covariant derivative.

\subsection*{Irreducibility}

We now show that the sets 
\be
\{e^{-i\omega t}\bm{\mathcal{S}}_{(+)}^{klm}\},\{e^{-i\omega t}\bm{\mathcal{S}}_{(-)}^{klm}\},\{e^{i\omega t}\bm{\mathcal{S}}_{(+)}^{klm}\},\{e^{i\omega t}\bm{\mathcal{S}}_{(-)}^{klm}\}
\label{sets}
\ee
do not mix under the action of the group. Since the SO(1,4) algebra is generated by the SO(4) generators and the boost generator 
$$\bm D = \sin t \cos\chi \,\bm \partial_t +\cos t \sin \chi\, \bm \partial_\chi\,,$$  
it is enough to show that the action of $\bm D=d^\mu\bm\partial_\mu$ is closed on each set. Since we are working with gauge fields instead of field strengths, we should take into account that the Lie derivative along $\bm D$  take the sets away from the temporal gauge \eqref{Tg}. We can fix this compensating the Lie derivative action with a gauge transformation. The appropriate gauge transformation can be read from 
$$\delta_{\bm D} A_t=\pounds _{\bm D}A_t=d^\mu\cancel{\partial_\mu A_t}+A_\mu \partial_t d^\mu=\partial_t\Big(\underbrace{\int^tdt'A_\chi \partial_{t'} d^\chi}_{\alpha(t,\Omega)}\Big)$$
here we used that $A_t=d^\theta=d^\phi=0$. Hence, if we subsequently perform the gauge transformation given by
$$\delta_\alpha A_\mu=-\partial_\mu\alpha$$
we return to the temporal gauge. 
Defining $\bar\delta_{\bm D}\equiv\delta_{\bm D}+\delta_\alpha$, one finds that the first set is closed under the action of $\bar\delta_{\bm D}$
$$\bar\delta_{\bm D}\big(e^{-i(k+1) t} \bm{\mathcal{S}}_{(+)}^{klm}\big)=\frac{k+l+2}2e^{-i(k+2) t} \bm{\mathcal{S}}_{(+)}^{(k+1)lm}-\frac{k-l}2e^{-i k  t} \bm{\mathcal{S}}_{(+)}^{(k-1)lm}$$
Notice that positive chiralities do not mix with negative ones. Also, since $1\le l\le k$ we cannot connect positive with negative energies. Similar equations follows for the rest of the sets in \eqref{sets}.

\subsection*{\bf Quantisation} 

The normalisation ${\cal N}_k$ in (\ref{canonical_modes})  is fixed in the usual way, so that the mode functions 
$$ A^{klm}_{(\sigma)a}=\mathcal{N}_k   e^{-i \omega t}  {\mathcal{S}}_{(\sigma)a}^{klm} (\Omega)$$ 
and its (negative energy) conjugates satisfy
\begin{equation}
    (A_n, A_{m}) = \delta_{nm} \, , \quad (A^*_n, A^*_{m}) = - \delta_{nm} \, , \quad (A_n, A^*_{m}) = 0 \, ,
\end{equation}
here we succinctly denoted $n=(\sigma klm)$. The de Sitter invariant  inner product is given by
\begin{align}
(A ,A')&=-i \int d^3\Omega\sqrt {-g} g^{\mu\nu}A_\mu^{*}(t,\Omega) \overset{\leftrightarrow}{D^t} A'_\nu(t,\Omega)\nn\\
&=i \int d^3\Omega\sqrt {\mathfrak g} \mathfrak g^{ab}A_a^{*}(t,\Omega) \overset{\leftrightarrow}{\partial_t} A'_b(t,\Omega)
\label{def:vector_sp}
\end{align}
where $dtd\Omega\sqrt{-g}$ is the spacetime volume element  and $D^\mu=g^{\mu\nu}D_\nu$ is the covariant derivative in de Sitter space. In passing to the second line we restricted to conformal global coordinates where the conformal factors and Christoffell symbols  cancel (see \eqref{ChdS}). We have also considered that the gauge field is in the temporal gauge.  The result for the normalisation is 
\begin{equation}
\mathcal{N}_k = \frac{1}{ \sqrt{2 \omega(k)}} \, . 
\label{res:norm}
\end{equation}
We can now proceed to canonically quantise the theory, 
$$ 
[\hat{A}_a(t,\Omega), \hat{A}_b(t,\Omega')] =   [\hat{\pi}_a(t,\Omega), \hat{\pi}_b(t,\Omega')] = 0 \, , 
$$ 
\begin{equation}
[\hat{A}_a(t,\Omega), \hat{\pi}_b(t,\Omega')] = i\Pi_{ab} \,\frac{\delta^{(3)}(\Omega-\Omega')}{\sqrt{\mathfrak{g}}}  \, ,
\label{CCR}
\end{equation}
where $\Pi_{ab}$ is a projector on the transverse polarizations. The conjugate momentum is defined as
\begin{equation}
\pi^a = \frac{\partial \sqrt {-g}\mathcal{L}}{\partial \dot{A}_a} =- \sqrt{-h} \,h^{0\mu}h^{a\nu} F_{\mu\nu} = \sqrt{\mathfrak g} \mathfrak g^{ab} \partial_t A_b \, .
\end{equation}
The canonical commutation relations \eqref{CCR} imply
\begin{equation}
[\hat{a}_ {n},\hat{a}_{n'}] =   [\hat{a}_{n}^ \dagger,\hat{a}_{n'}^ \dagger ] = 0 \, , \quad [\hat{a}_n,\hat{a}_{n'}^\dagger] =  \delta_{nn'} \, .
\end{equation}
By explicit computation we obtain 
\begin{equation}
    [\hat{A}_a(t,\Omega), \hat{\pi}_b(t,\Omega')] = i \sum_{k,l,m} \sum_{\sigma = \pm} \left( \mathcal{S}_{(\sigma)a}^{klm}(\Omega) \mathcal{S}_{(\sigma)b}^{\dagger klm}(\Omega') \right) \, ,
\end{equation}
Comparing with \eqref{CCR} and using the completeness relation \eqref{CRel} we find the expected result 
\begin{equation}
    \Pi_{ab} = \mathfrak{g}_{ab} - \frac{\nabla_a \nabla'_b}{-\nabla^{'2}_{(0)}}  \, . 
\end{equation}
here $\nabla'$ denotes derivative with respect to $\Omega'$.

\vspace{2mm}

\nin Notice that the Euclidean continuation of \eqref{def:ds_conformalmetric}, under $t\to \tfrac \pi2+iX$, becomes the round sphere. Since the positive energy modes in (\ref{canonical_modes}) become regular on the south pole ($X\to-\infty$) in the Euclidean section, we conclude that the vacuum defined as
\begin{equation}
    \hat a_{(\sigma )}^{klm}\lvert BD \rangle = 0 \,,\qquad\forall ~ (\sigma) {klm}\,.
\end{equation}
corresponds to the Euclidean/Bunch-Davies vacuum state  \citep{Higuchi:2010xt, Schlingemann:1999mk}.

\subsection*{Gauge invariant operators} 

The gauge invariant Electric $\hat{\bm{  {E}}}$ and Magnetic fields ${\hat{\bm{  {B}}}}$ follow from $\hat F_{\mu\nu}$.  The spatial components ${F}_{ab}$ read 
\begin{equation}
\hat{F}_{ab}(t,\Omega) = \sum_{\sigma=\pm} \sum_{klm} \mathcal{N}_k \left( e^{-i \omega t} \nabla_{[a} \mathcal{S}_{(\sigma)b]}^{klm} (\Omega)\, \hat a_{(\sigma )}^{klm} + e^{i\omega t}\nabla_{[a}\mathcal{S}_{(\sigma)b]}^{\dagger\, klm} (\Omega) \, \hat{a}^{\dagger\,{klm}}_{(\sigma)}  \right) \, .
\label{Ffinal}
\end{equation}
From which we obtain the Magnetic field as
\begin{equation}
\hat{ {B}}_a(t,\Omega) = \frac12\frac{\epsilon_a{}^{bc} }{\sqrt{\mathfrak g}}\hat{F}_{bc} \, .
    \label{magneticF}
\end{equation}
The Electric field is given by
\begin{equation}
    \hat{ {E}}_a(t,\Omega) =- \partial_t \hat{A}_a(t,\Omega)\,.
\end{equation}
From \eqref{EiBjuntitos} we recognise that, naturally, on the solutions of the equations of motion $\hat{\bm{  {E}}}$ and   ${\hat{\bm{  {B}}}}$ are not independent. Taking the magnetic field components as the independent degrees of freedom, the discussion of the previous section implies that its chiral components  
\begin{equation}
    \hat{ {B}}_a(t,\Omega) = \hat{\mathcal{B}}^+_a(t,\Omega) + \hat{\mathcal{B}}^-_a(t,\Omega)  \, .
\end{equation}
acting on the Bunch-Davies vacuum state span two Discrete Series modules 
\begin{equation}
   \hat{\mathcal{B}}_a^+ \lvert BD \rangle + \hat{\mathcal{B}}_a^- \lvert BD \rangle \simeq \mathcal{U}_{1,0}^+ \oplus \mathcal{U}_{1,0}^- \, .
\end{equation}
The quantum number $\bm s = 1$ follow from the vector character under SO$(3)$, while the $\Delta = 2$ scaling dimension  can be seen either from the fact that the SO$(1,4)$ Casimir \eqref{Cvec} vanishes (cf. \eqref{AdSbox}) or from the late time expansion of the field \citep{Sengor:2019mbz, Sengor:2022hfx, Sengor:2022kji}.  

For completeness we mention that the SO$(4)$ content of $\hat F_{\mu\nu}$ is immediate and  can be inferred from  the quantum numbers of the mode functions. As discussed in  detail in the App. \ref{eigenD}, representations of SO$(4)$ are labelled by the two Casimirs (\ref{CasimirsMain}). Positive energy mode function therefore give rise to
\begin{equation}
    \hat F_{\mu\nu}\lvert BD \rangle  \simeq  \bigoplus_{k=1}^\infty  \mathbb{Y}_{k,+1}  \,  \oplus \,  \bigoplus_{k=1}^\infty \mathbb{Y}_{k,- 1}  
\end{equation}
where the notation for Young diagrams can be found in App. \ref{youngs}. Alternatively, we  can express the content of the single particle Hilbert space in terms of the SO$(4) \simeq \textnormal{SU}(2)_L \times \textnormal{SU}(2)_R$ decomposition 
\begin{equation}
    \hat F_{\mu\nu}\lvert BD \rangle  \simeq\bigoplus_{k=1}^\infty  \left[ \frac{k+1}{2}, \frac{k-1}{2} \right] \oplus\bigoplus_{k=1}^\infty  \left[\frac{k-1}{2}, \frac{k+1}{2} \right]  \, .
\end{equation}
here $[j_L, j_R]$ refer to the $\textnormal{SU}(2)_L \times \textnormal{SU}(2)_R$ quantum numbers.

\section{Summary \& Outlook}
In this note we studied the field theoretic realisation of the Discrete Series representation of SO$(1,4)$ in terms of a free spin$-1$ gauge field propagating in a fixed de Sitter background.  We revisited the Euclidean path integral quantisation and computed the $1-$loop Sphere partition function showing how it can be understood in terms of the Harish-Chandra character of SO$(1,4)$, and in general of SO$(1,d+1)$, UIRs. In addition to the character of the spin$-1$ Discrete Series UIR we showed how the \emph{edge mode} character appears. This contributions can be understood as  localised in a co-dimension $2$-surface to be associated to the presence of the cosmological horizon. For the case at hand, we showed that the \emph{edge modes} of the spin$-1$ Discrete Series correspond to a Discrete Series representation of SO$(1,2)$. 

We then solved the quantum gauge theory in real time showing the single particle Hilbert space of the theory  furnished the Discrete Series UIR. To this end, we canonically quantise the theory putting special focus in its group theoretic properties. We elucidate the role of the second quadratic Casimir of SO$(4)$ and showed how it allows one to construct the corresponding mode functions with the appropriate quantum numbers. This allowed us to identify two Discrete Modules, $\mathcal{U}_{1,0}^\pm$, of SO$(1,4)$ in the single particle Hilbert space of the theory. For the de Sitter global coordinate system, the \emph{edge modes} that appeared in the Sphere computation were not found to be present in the modules spanned by the quantum field strength.

The prefactor \eqref{prefactor} depending on $\textnormal{g}$  might come as a surprise. From a Lorentzian perspective the path integral should be counting states on the Hilbert space of the theory and as such this factor spoils this interpretation by being in general non-integer valued. Yet, this is a general feature of quantum fields in a compact manifold \citep{Donnelly:2013tia,Anninos:2020hfj,Giombi:2015haa,Witten:1995gf}. Even more, this was seen in Chern-Simons theory by different methods \citep{Witten:1988hf} and recently understood in the context of $3-$dimensional de Sitter spacetimes \citep{Anninos:2021ihe} where it was understood as stemming from a topological entanglement entropy computation \citep{Kitaev:2005dm}. \\

\nin We would like to conclude this note with some comments and outlook regarding the features that the Discrete Series representation of SO$(1,d+1)$ showcases and how we can further understand them.

\subsection*{Weakly interacting Einstein-Maxwell}
The lack of clear, diffeomorphism and field-redefinition invariant, \emph{observables} for the $\Lambda > 0$ Euclidean low-energy effective field theory makes the definition of coupling constants subtle. For a $4-$dimensional theory the most general effective action up to order $R^2$ involves also the Weyl tensor $C_{\mu \nu \rho \sigma}$ and two corresponding coupling constants $\lambda_{C^2}$, $\lambda_{R^2}$. As discussed before, the Euclidean path integral effectively defines a gauge invariant, diffeomorphism and field-redefinition invariant quantity $S_{dS}$ (\ref{dSentr}) that allow us to define the dimensionless coupling $(G \Lambda^{(d-1)/2})^{-1}$ in terms of the on-shell action. This gives us a way out, if the Euclidean effective action allows for different saddles $g_M$ we can use $S^{(0)}_M = - S_E[g_M]$ as defining the corresponding coupling constants. 

In $4-$dimensional theories the topologies known to admit $\Lambda > 0$ Einstein metrics are $S^4, S^2\times S^2, \mathbb{CP}^2$ and the connected sum $\mathbb{CP}^2\# k \mathbb{CP}^2$ with $1\leq k \leq 8$. Following general considerations \citep{Anninos:2020hfj} for a saddle $M$ we can repeat the exercise of Section \ref{secPI} and expand the UV-divergent terms as 
\begin{equation}
    \mathcal{S}_M = S^{(0)}_M - \frac{D_M}{2} \log S^{(0)}_M + \alpha_M \log \frac{\ell_0}{L} + K_M + \cdots \, ,
    \label{outlook1}
\end{equation}
where $S_M^{(0)}$ is the on-shell action, $D_M$ is the number of Killing vectors of $M$, and $\ell_0 = (3 \Lambda^{-1})^{1/2}$ and $K_M$ is a numerical constant obtained by evaluating (\ref{PI_FINAL}). The existence of the different saddles allow us to define meaningful couplings $\Lambda_{C^2}, \lambda_{R^2}$ through different combinations of (\ref{outlook1}). 

If we now consider the inclusion of a spin$-1$ gauge field we need to properly define one more coupling, namely $\textnormal{g}_{YM}$. Computing (\ref{outlook1}) for $S^4, S^2 \times S^2, \mathbb{CP}^2$ allows us to define unambiguously the $2$ gravitational couplings $\lambda_{C^2}, \lambda_{R^2}$. To also define $\textnormal{g}_{YM}$ we need to compute (\ref{outlook1}) on $\mathbb{CP}^2\#k \mathbb{CP}^2$. One would then be lead to question, is it possible to define an infinite number of coupling constants in dS?. Contrary to the S-matrix case in which we have the S-matrix to provide and infinite number of gauge invariant information to define them, here we just have a finite set of gauge invariant information to define them. Thus we are left with the question of how do we properly define the couplings of a family of interacting particles in dS spacetime? Already for different families of gauge fields as the one we found in our universe the answer is not trivial. For $k \geq 5$ the connected sum $\mathbb{CP}^2\#k \mathbb{CP}^2$ develops a moduli space of nonzero dimension \citep{moduli1,moduli2}. We expect that coupling gravity in dS with different species of gauge fields allows us to probe this moduli space and find the proper way to define the corresponding gauge couplings. These issues will be discussed somewhere else.

\subsection*{Breathing life into the Edge Modes}
The $1-$loop sphere path integral encodes Lorentzian information in the form of Harish-Chandra characters of the corresponding UIR. In addition to the bulk degrees of freedom (\ref{bulk_character_PI}) there are \emph{edge modes} (\ref{edge_character_PI}) corresponding to a $2-$dimensional Discrete Series UIR. These UIRs have been shown to arise in the free theory, thus we want to further understand their dynamics and whether they interact with the bulk degrees of freedom. 

We could envision considering an interacting gauge theory.\footnote{For example a weakly interacting Einstein-Maxwell theory.} The $2-$point structure of the interacting $2-$point function can be studied with the  
Källén-Lehmann spectral decomposition \citep{Kallen:1952zz, Lehmann:1954xi,Loparco:2023rug,Hollands:2011we,DiPietro:2021sjt,Bros:1990cu,Bros:1995js,Bros:2009bz}
\begin{equation}
    \langle \mathcal{O}(x) \mathcal{O}(x') \rangle_I = \int_\mathcal{C} \frac{\textnormal{d}\Delta}{2\pi i}  \rho(\Delta) \langle \mathcal{O}(x) \mathcal{O}(x') \rangle_{\textnormal{Free}}
\end{equation}
where the contour $\mathcal{C}$ is around the values of $\Delta$ defining the corresponding UIR, $\rho(\Delta)$ is a non-negative spectral function and $\langle \mathcal{O}(x) \mathcal{O}(x') \rangle_{\textnormal{Free}}$ is the corresponding free $2-$point function.\footnote{Generalisations to operator with space-time indices is straightforward.}  

The free $2-$point function can readily be computed from the partition function upon introducing a suitable source term for the field $\mathcal{O}$. Furthermore, since now the theory is free this $2-$point function will in turn be written in terms of (\ref{PI_FINAL}). Thus characterising the analytic properties of the spectral function $\rho(\Delta)$ would characterise how, or if, the \emph{edge modes} interact with the bulk degrees of freedom. 

For the spin$-1$ field we now know more about the \emph{edge modes}. They correspond to a SO$(1,2)$ Discrete Series representation with $\Delta = 1$ yielding the unitary character
\begin{equation*}
    \chi_{\mathcal{D}_\Delta^+ \oplus \mathcal{D}_\Delta^-} (t) = 2 \frac{q}{1-q} \, ,
\end{equation*}
a field theory realisation devoid of problems was recently constructed as a theory of a massless scalar field with the shift symmetry gauged  \citep{Anninos:2023exn}. An in depth study of such theory, its sphere path integral and the connection to BF-theories will also shed light into the physics of the \emph{edge modes} of the U$(1)$ gauge field. 

Furthermore, any gauge theory has also in its spectrum non-local operators such as Wilson lines. Such operators might play a role in describing the \emph{edge modes} of the theory upon cutting the geometry with the cosmological horizon. This has been discussed in the context of lattice gauge theories \citep{Buividovich:2008kq}, understanding this for $4-$dimensional de Sitter spacetime is another future venue of research.

\subsection*{Cornucopia of Discrete Series}
The Discrete Series representation is ubiquitous to the SO$(1,d+1)$ representation theory. From a physical point of view, in general, they correspond to spin$-s$ gauge fields. Furthermore, there are also fermionic counterparts of these representations whose properties have yet to be fully understood \citep{ottoson,schwarz,Basile:2016aen}. For the $4-$dimensional case the simplest example corresponds to a spin $\bm s = \frac{3}{2}$ field \citep{Letsios:2023qzq}. A field theory realisation of this fermionic Discrete Series UIR requires an imaginary \emph{mass}, rendering the unitarity of the theory subtle. Following the discussion of the note, a proper $1-$loop calculation of the $\bm s = \frac{3}{2}$ field should yield the Harish-Chandra character of the corresponding UIR, showing that the tension stemming from the imaginary \emph{mass} parameter is an artifact of the Lagrangian formulation \citep{nosotros}. 

Furthermore, there are also higher spin fermionic generalisation of the Discrete Series representation ($\bm s = \frac{5}{2}, \frac{7}{2}, \cdots$) and $p-$form analogues \citep{David:2021wrw}. Bosonic higher spin fields give one of the most concrete microscopic realisations of de Sitter quantum gravity \citep{Anninos:2017eib}. Properly characterising the fermionic higher spin fields will allow to generalise this construction.

\section*{Acknowledgements}

We thank D. Anninos, D. Galante V. Letsios and S. Vitouladitis for interesting discussions on this and related topics and feedback on the manuscript of this paper. We thank also A. Higuchi for correspondence and sharing with us his PhD Thesis. GAS would like to thank King's College, London and the Royal Society for hospitality and financial support.

This work was funded by CONICET grants  PIP-UE$ 084$ and UNLP grant $X791$ and  PICT 2020-03826. MS is supported by a CONICET fellowship. The work of ARF was supported by the Royal society grant RF/ERE/210168 which is part of the Royal Society URF grant “The Atoms of a de Sitter Universe”.

\appendix

\section{UIRs of SO$(1,d+1)$}
\label{app:group}
In this appendix we spell out the details and conventions we use to characterise irreducible representations of  SO$(1,d+1)$. 

\subsection{de Sitter isometry and Euclidean conformal group} 
\label{dsCFT}

The isomorphism between the de Sitter generators (\ref{def:ds_algebra}) and the $d$-dimensional Euclidean conformal group follows  from the identification\footnote{Embedding space indices are denoted $A=(0,i,d+1)$ with $i,j= 1,2,...\,, d $ the CFT indices.}   

\vspace{-4mm}
\begin{equation}
L_{ij} = M_{ij} \, , \quad L_{0,d+1}=D \, , \quad L_{d+1,i} = \frac12 (P_i + K_i) \, , \quad L_{0,i} = \frac12 (P_i - K_i)\,.
\label{def:conformal_iso}
\end{equation}
Here $D$ is the dilatation operator, $P_i$  correspond to translations, $K_i$ give special conformal transformations and $M_{ij}=-M_{ji}$ are  rotation generators in $\mathbb R^d$. The commutation relations (\ref{def:ds_algebra})  now  take the form
\begin{equation}
\begin{split}
[D,P_i] &= P_i \,, \quad [D,K_i] = - K_i \, , \quad [K_i,P_j] = 2\delta_{ij}D-2 M_{ij} \, , \\ 
[M_{ij},P_k] &= \delta_{jk}P_i - \delta_{ik}P_j \, , \quad [M_{ij},K_k] = \delta_{jk} K_i - \delta_{ik} K_j \, ,\\ 
[M_{ij},M_{kl}] &= \delta_{jk} M_{il} - \delta_{ik} M_{jl} + \delta_{il} M_{jk} - \delta_{jl} M_{ik} \, .
\label{def:conf_algebra}
\end{split}
\end{equation}
The generators $L_{AB}$ exponentiate to give SO$(1,d+1)$ group elements. Near the identity we have   
\begin{equation}
g(\theta) = \exp\left(\frac{1}{2} \theta^{AB} L_{AB} \right) \, , 
\end{equation}
with $\theta^{AB} = - \theta^{BA}$  real parameters. 

\nin There are several subgroups  within SO$(1,d+1)$. They are 
\begin{equation}
	\begin{split}
	\textnormal{K} &= \text{SO}(d+1)\, , \quad \textnormal{M} = \{ e^{\frac{1}{2}\omega^{ij}M_{ij}}, \, \omega^{ij} \in \mathbb{R} \} = \text{SO}(d) \, , \\ 
	\textnormal{N} &= \{ e^{b\cdot K}\, , b^i \in \mathbb{R} \}\, , \quad \tilde{\textnormal{N}}= \{ e^{x\cdot P}\, , x^ i \in \mathbb{R} \} \, , \quad \textnormal{A} = \{ e^{\lambda D}\, , \lambda \in \mathbb{R} \} \, ,
	\end{split}
	\label{def:so_subgroups}
\end{equation}
K is the maximal compact subgroup of SO$(1,d+1)$ and corresponds to the isometry group of the constant time slices, namely $S^d$, in global coordinates. Notice that K differs from M, i.e. the rotation group of $\mathbb R^d$. The  construction of the SO$(1,d+1)$ UIR's is  similar to what we do for the Poincare group. UIRs are build  from irreps  of the stability group of $SO(1,d+1)$, this happens to be the composition the NAM subgroups \citep{Dobrev:1977qv}.\footnote{For the Poincare group, the stability group is defined as the subgroup H $\subset$ ISO(1,d) that leaves invariant a given reference momentum. It is traditionally called little group.} If we realise SO($1,d+1$) as the conformal group of $\mathbb R^d$, the stability group corresponds to the transformations that leave invariant the origin of $\mathbb R^d$.

From (\ref{def:conf_algebra}),  the quadratic Casimir \eqref{Casimir_sec1} reads
\begin{equation}
\mathcal{C}_2 = \frac{1}{2} L_{AB}L^{AB} = -D(D-d) + P_i K_i + \frac{1}{2} M_{ij}^ 2  \, .
\label{def:casimir_explicit}
\end{equation}
Here, $\frac12 M_{ij}^2$ is  the quadratic Casimir of the subgroup M. As an example, for $d=3$ we find $\frac12 M_{ij}^2=-s(s+1)$.

\subsection{Classification of UIRs}
\label{youngs}

Unitary (infinite dimensional) representations of SO$(1,d+1)$ are labelled in terms of  labels for SO$(1,1)\times\,$SO$(d)\subset\; $SO$(1,d+1)$. These are the scaling dimension $\Delta \in \mathbb{C}$ for SO$(1,1)$ and a highest-weight vector $\bm{s}=(s_1,s_2,\cdots, s_r)$  for SO$(d)$. Recall that for SO($d$)-groups the number of components of $\bm s$ is $r = \left[  \frac{d}{2} \right]$. The  components of the highest weight vector $\bm s$ follow  the convention
\begin{equation}
s_1 \geq s_2 \geq \cdots \geq s_{r-1}\geq \lvert s_r \lvert \, .
\end{equation} 
with $s_i$ being all positive integer (or all half-integers) except for the last component $s_r$ which is always positive for odd dimensional SO$(2r+1)$, but can be either positive or negative for even dimensional SO$(2r)$. This sign is called `chirality' of the representation. To any given highest weight vector $\bm s$ we associate a Young diagram $\mathbb Y_{\bm s}$ in the usual way, it consisting of $r$-rows such that there are $s_i$
boxes in the $i$-th row (in the case of negative $s_r$ we ignore the sign). In the following the relevant  highest weight vectors are $\bm s=(s,0,0,...0)$ and $\bm s=(n,m,0...0)$ will appear,  their Young diagrams will be denoted $\mathbb Y_s$ and $\mathbb Y_{n,m}$ respectively.

Single-row Young diagrams, i.e. $\bm s  = (s,0,\cdots,0)$, labelled by a non-negative integer $s$ map to symmetric traceless tensors of  SO$(d)$ with $s$-indices on dS$_{d+1}$. As customary we will refer to them as the spin-$s$ representation (see \citep{Sun:2021thf} for a review). More general highest-weight vectors $\bm{s}$ correspond to fields of mixed symmetry \citep{Basile:2016aen,Letsios:2022slc,Pethybridge:2021rwf,Letsios:2020twa}. For any $d \geq 3$, there are four types of SO$(1,d+1)$  UIRs apart from the trivial representation \citep{Dobrev:1977qv,Basile:2016aen,Sun:2021thf}. Their decomposition into the maximal compact subgroup SO($d+1$) is
\begin{itemize}
\item{\textbf{Principal series} $\mathcal{P}_{\Delta,s}$: $\Delta \in \frac{d}{2} + i \mathbb{R}$ and $s \geq 0$. The restriction of $\mathcal{P}_{\Delta,s}$ to the maximal compact subgroup K of (\ref{def:so_subgroups}) is given by 
\begin{equation}
\mathcal{P}_{\Delta,s}\Big\rfloor_{SO(d+1)} = \bigoplus_{n=s}^{\infty} \bigoplus_{m=0}^s \mathbb{Y}_{n,m} \, , 
\end{equation}
here $\mathbb{Y}_{m,n}$ denotes a two-row Young diagram with $n$ boxes in the first row and $m$ boxes in the second row.}
\item{\textbf{Complementary series} $\mathcal{C}_{\Delta,s}$: $0 < \Delta < d$ when $s = 0$ and $a < \Delta < d-1$ when $s \geq 1$. It has the same $SO(d+1)$ contents as $\mathcal{P}_{\Delta,s}$}
\item{\textbf{Type I exceptional series} $\mathcal{V}_{p,0}$: $\Delta = d+p-1$ and $s = 0$ for $p \geq 1$. The $SO(d+1)$ content of $\mathcal{V}_{p,0}$ only consists of single-row Young diagrams: 
\begin{equation}
\mathcal{V}_{p,0}\Big\lvert_{SO(d+1)} = \bigoplus_{n=p}^\infty \mathbb{Y}_n
	\end{equation}}
	\item{\textbf{Type II exceptional series} $\mathcal{U}_{s,t}$: $\Delta = d + t - 1$ and $s \geq 1$ with $t = 0, 1, 2, \cdots, s-1$. The $SO(d+1)$ content is
	\begin{equation}
		\mathcal{U}_{s,t}\Big\lvert_{SO(d+1)} = \bigoplus_{n=s}^\infty \bigoplus_{m=t+1}^s \mathbb{Y}_{n,m} \, .
	\end{equation}}
	\label{def:irreps_boson}
\end{itemize}
The representations labeled by $[\Delta,s]$ and $[d-\Delta,s]$ in the principal and complementary series are actually isomorphic, therefore we only consider $\Delta$ with non-negative imaginary part in principal series, and $\Delta > \frac{d}{2}$ in complementary series. Given a scaling dimension $\Delta$ we use the notation $\mathcal{F}_{\Delta, s}$ for both the principal series or complementary series, each UIR (\ref{def:irreps_boson}) is in a one to one correspondence with a free field in $dS_{d+1}$
\begin{itemize}
\item{$\mathcal{F}_{\Delta,s}$: describes spin$-s$ massive fields of mass
\begin{equation}
\begin{split}
m^2 &= \Delta(d-\Delta) \, , \qquad \textnormal{for} \quad s = 0 \\
m^2 &= (\Delta + s - 2)(d+s-2-\Delta) \, , \qquad \textnormal{for} \quad s \geq 1
\end{split}
\end{equation}}
\item{$\mathcal{U}_{s,t}$: describes partially massless gauge fields of spin $s$ and depth $t$. For $t = s-1$ the field corresponds to an exactly massless gauge field, \emph{i.e} photon, graviton, etc.}
\item{$\mathcal{V}_{p,0}$: is expected to describe scalar fields of mass 
\begin{equation}
m^2 = (1-p)(d+p-1) \, ,
\end{equation}
with some shift symmetry being gauged.}
\end{itemize}

\section{Eigenfunctions on $S^D$}
\label{eigenD}

In this appendix we collect details for the characterisation of the eigenmodes of symmetric traceless tensor fields of spin-$s$ on $S^{d+1}$. For a full derivation we refer the reader to the classic paper \citep{Higuchi:1986py}. The  $S^3$ solutions needed in the bulk of the text to quantise  the Maxwell field are discussed in full detail in the next subsection.  

\subsection{Eigenfunctions}
A symmetric traceless spherical harmonic (STSH's) defined on $S^{d+1}$ is given by the symmetric tensor eigenfunctions $h_{\mu_1 \mu_2 \cdots \mu_s}$ of the Laplace operator obeying 
\begin{equation}
    \begin{split}
        -\nabla^2 h_{\mu_1 \mu_2 \cdots \mu_s} &= \lambda h_{\mu_1 \mu_2 \cdots \mu_s} \\ 
        \nabla^\mu h_{\mu \mu_1 \cdots \mu_{s-1}} &= 0 \\ 
        g^{\mu \nu} h_{\mu \nu \cdots \mu_{s-2}} &= 0 \, , 
    \end{split}
\end{equation}
where $g^{\mu \nu}$ is the inverse metric on $S^{d+1}$. These functions admit an expansion in a complete basis of STSH that we define by $f_{n,\mu \nu \cdots \rho_s} = f_{n,(s)}$ and satisfy 
\begin{equation}
    -\nabla^2_{(s)} f_{n,(s)} = \lambda_{n,s} f_{n,(s)}\,, \quad \nabla \cdot f_{n,(s)} = 0\, , \quad f^\alpha_{n,\alpha, (s-1)} = 0 
    \label{app:stsh_modes}
\end{equation}
with eigenvalues and degeneracies given by 
\begin{equation}
    \begin{split}
        \lambda_{n,s} &= n(n+d) - s\, ,\quad n \geq s \\ 
        D_{n,s}^{d+2} &= g_s \frac{(n-s+1)(n+s+d-1)(2n+d)(n+d-2)!}{d!(n+1)!} \, , \\ 
        g_s &= \frac{(2s+d-2)(s+d-3)!}{(d-2)!s!} \, .
    \end{split}
    \label{eigenvalues}
\end{equation}
Defined like this, these eigenfunctions furnish irreducible representations of SO$(d+2)$ corresponding to two-row Young diagrams with $n$ boxes in the first row and $s$ boxes in the second row. They are normalised with respect to the inner product 
\begin{equation}
    (h_{(s)}, h'_{(s)}) = \int_{S^{d+1}} d^{d+1}x \sqrt{g} h^{\mu_1 \cdots \mu_s} h'_{\mu_1 \cdots \mu_s} \, ,
\end{equation}
defined like this (\ref{app:stsh_modes}) the eigenmodes satisfy 
\begin{equation}
    (f_{n,(s)}, f_{m,(s)}) = \delta_{nm} \, , 
\end{equation}
there is also a specific set of modes that correspond to the case $n = s$ that obey 
\begin{equation}
    \nabla_{(\mu_1} \varepsilon_{\mu_2 \cdots \mu_{s+1})} = 0 \, ,
\end{equation}
it can be seen that defined like this it is a spin$-s$ Killing tensor.

\subsection{Scalar and Vector  Harmonics on $S^3$ }
\label{S3irreps}

In this section we consider in detail the 3-sphere case, spelling out the explicit formul\ae{} for the scalar and vector harmonics. Special attention is paid to the relation between $S^3$ and the $SU(2)$ group manifold. Group theoretic properties of the construction  play an important  role in the representation theory of the gauge field in d$S_4$. 

The isometry group of $S^3$ is given by SO$(4)$ whose anti-Hermitian generators  satisfy ($  K_{ij}=-  K_{ji}$)
\begin{equation}
    [K_{ij}, K_{kl}] = -  \delta_{ik} K_{jl} - \delta_{jl} K_{ij} + \delta_{il} K_{jk} + \delta_{jk} K_{il}   ,~~~~  (i,j=1,..4 )
\end{equation}
The SO(4) algebra posses two  Casimirs, they are given by
\begin{equation}
    \mathcal{C}  = \frac{1}{8} K_{ij}K^{ij} \, , \qquad \tilde{\mathcal{C}}  = \frac{1}{32} \epsilon^{ijkl} K_{ij}K_{kl} \, ,
    \label{casimir_gen}
\end{equation}
where $\epsilon^{ijkl}$ is the $4$-index Levi-Civita symbol. As well known, SO(4) $\sim$ SO(3) $\times $ SO(3). This fact is easily recognised defining 
\begin{alignat}{2}
    {  L}_1 &=  K_{14}+K_{23}  \, , \qquad {  R}_1 &&= K_{14}-K_{23}  \, , \\
      L_2 &=  K_{24}+K_{31}  \, , \qquad {  R}_2 &&= K_{24}-K_{31} \, ,\\ 
    {  L}_3 &=  K_{34}+K_{12}  \, , \qquad {  R}_3 &&=  K_{34}-K_{12} \, ,
\end{alignat}
Then,
\be
[L_i,L_j]=-2\epsilon_{ijk}L_k,~~~[R_i,R_j]=2\epsilon_{ijk}R_k,~~~[L_i,R_k]=0
\label{LRal}
\ee
showing that $\{L_i\}$ and $\{R_i\}$ give two commuting $\mathfrak{su}(2)$ algebras. This result tells us that unitary irreducible representations of SO(4) will be labelled by two SU(2) quantum numbers, namely 
$$\text{SO(4) \sf labels}:~~~[ j_L,j_R]$$
The Casimirs \eqref{casimir_gen} in the $\{L_i,R_i\}$-basis take the form
\begin{align}
&{\cal C} =\frac12\left[  (L_i)^2+(R_i)^2\right],&&\tilde{\cal C} =\frac14\left[ (L_i)^2-(R_i)^2\right]\nn\\
&~~=-2\left[j_L(j_L+1)+j_R(j_R+1)\right],&&~~=-j_L(j_L+1)+j_R(j_R+1)
\label{su2x}
\end{align}
On the other hand, it is well known that $S^3$ is the group manifold of SU$(2)$ and that we can  obtain the SO(4)  Killing vectors from the bi-invariant metric on SU(2). We summarise below the key concepts relating  group theory to geometry. For a general introduction see \citep{GS,gibbDAMTP}, and \citep{Kumar:2020xjr} for related work. 

\vspace{2mm}

\subsection*{ $S^3$ geometry and the SU(2) group}
Start by parametrising an SU$(2)$ element as
\begin{align}
U(\chi,\theta,\phi)&=\exp(  i \chi\: ({\bm n}\cdot\bm \sigma))
\label{U}
\end{align}
with ${\bm n}=(\sin\theta\cos\phi,\sin\theta \sin\phi ,\cos\theta )$ a unit vector on $\mathbb R^3$, and $\sigma^i$ the Pauli matrices. The SU$(2)_L$ left invariant 1-forms $\bm \lambda$ are defined as
\begin{equation}
    \bm \lambda  := -i U^{-1}\bm dU  \, ,
    \label{b1}
\end{equation}
they are invariant under $L$-translations $U(\chi,\theta,\phi)\mapsto V_L\, U(\chi,\theta,\phi)$ for constant $V_L$.
These, allow us to obtain the $L$-invariant metric as
\begin{equation}
    \textnormal{d}s^2 = \frac{1}{2} {\sf tr}[\, \bm \lambda\,  \bm \lambda] =  d\chi ^2+\sin^2\chi(d\theta^2+\sin^2\theta  d\phi ^2) \, ,
    \label{binv}
\end{equation}
which we recognise as the round metric on  $S^3$. Here $ \chi,\theta\in(0, \pi)$ and $ \phi \in(0,2\pi)$.  The Ricci and scalar curvature for the round $S^3$ metric $\mathfrak g_{ab}$ are
\be
R_{ab}[\mathfrak g]=2{\mathfrak g}_{ab},~~~R[\mathfrak g]=6.
\label{s3g}
\ee
The enhancement from SU$(2)_L\to \textnormal{SO}(4)\sim \textnormal{SU}(2)_L\times \textnormal{SU}(2)_R$ arises from the fact that the $R$-action $U(\chi,\theta,\phi)\mapsto U(\chi,\theta,\phi)\,V_R$ is also a symmetry of \eqref{b1}. In other words,  \eqref{binv} is bi-invariant under  
$$U(\chi,\theta,\phi)\to V_L\, U(\chi,\theta,\phi)\,V_R \, ,$$ 
for  $V_{L},V_R$ independent constant $SU(2)$ matrices. \\
Writing the Maurer-Cartan 1-forms as $\bm\lambda= \bm\lambda^i\sigma^i$ we have
\begin{align}
\bm\lambda^1=\sin\theta \cos\phi \,\bm d\chi&+  (\frac12\sin2\chi\cos\theta\cos\phi-\sin^2\!\chi\sin\phi )\bm d\theta\nn \\
&-\frac12   (\sin^2\!\chi \sin2\theta \cos\phi + \sin2\chi\sin\theta\sin\phi)\bm d\phi\nn \, ,\\  
\bm\lambda^2=\sin\theta \sin\phi \,\bm d\chi&+ (\sin^2\!\chi\cos\phi+\frac12\sin2\chi\cos\theta\sin\phi)\bm d\theta\nn \\
&+\frac12   (\sin2\chi \sin\theta\cos\phi-\sin^2\!\chi \sin2\theta \sin\phi )\bm d\phi\nn \, ,\\
\bm \lambda^3=\cos\theta \,\bm d\chi-\frac12 &\sin2  \chi\sin\theta  \,\bm d\theta +   \sin^2\!\chi \,\sin^2\theta \bm d\phi\nn \, .
\end{align}
The $\mathfrak{su}(2)_L$ Killing vectors $\bm L_j$ of the metric \eqref{binv} are found to be duals to these 1-forms, i.e.   $\langle\bm\lambda^i|\bm L_j\rangle=\delta^i_j$ ($i,j=1,2,3$). They read
\begin{align}
\bm L_1=&\sin\theta \cos\phi \,\bm \partial_\chi+( \cot\chi\cos\theta\cos\phi-\sin \phi)\bm \partial_\theta -  (  \cot\chi \csc \theta\sin\phi+\cot\theta \cos\phi )\bm \partial_\phi\nn \, ,\\
\bm L_2=&\sin\theta \sin\phi \,\bm \partial_\chi+ (\cot\chi\cos\theta\sin\phi+  \cos\phi)\bm \partial_\theta+    (\cot\chi \csc\theta\cos\phi-\cot\theta \sin\phi)\bm \partial_\phi\nn \, ,\\
\bm L_3=&\cos\theta \,\bm \partial_\chi- \cot \chi\sin\theta \, \bm \partial_\theta +\bm \partial_\phi
\label{LIVF} \, .
\end{align}
One can verify the $L$-vectors and $L$-invariant forms satisfy
$$(\bm L_i)^a=\mathfrak g^{ab}(\bm \lambda^i)_b$$
In analogous way we define the right  invariant forms
$$\bm\rho:=-i\,\bm dU\,U^{-1}$$
these are invariant under $R$-translations $U\mapsto UV_R$ with $V_R= const$.
Writing $\bm\rho= \bm\rho^i\sigma^i$, one obtains,  
\begin{align}
\bm \rho^1=\sin\theta \cos\phi \,\bm d\chi+&(\sin^2\!\chi\sin\phi+\tfrac12\sin2\chi  \cos\theta\cos\phi)\bm d\theta\nn\\
&+ \tfrac12  (\sin^2\!\chi \sin2 \theta\cos\phi-\sin2\chi\sin\theta \sin\phi)\bm d\phi\nn \, ,\\
\bm \rho^2=\sin\theta \sin\phi \,\bm d\chi+& (\tfrac12 \sin2\chi\cos\theta \sin\phi-\sin^2\!\chi\cos \phi)\bm d\theta \nn\\
&+\tfrac12(\sin2\chi\sin\theta \cos\phi + \sin^2\!\chi\sin2\theta\sin\phi)\bm d\phi\nn \, ,\\
\bm \rho^3=\cos\theta \,\bm d\chi-\frac12\sin&2 \chi  \sin\theta \, \bm d\theta -\sin^2\!\chi\sin^2\!\theta\,\bm d\phi\nn \, .
\end{align}
The $\mathfrak {su}(2)_R$ Killing vectors are then given by
\begin{align}
\bm R_1=&\sin\theta \cos\phi \,\bm \partial_\chi+( \cot\chi\cos\theta\cos\phi+\sin \phi)\bm \partial_\theta +   (\cot\theta \cos\phi-  \cot\chi \csc \theta\sin\phi)\bm \partial_\phi\nn \, ,\\
\bm R_2=&\sin\theta \sin\phi \,\bm \partial_\chi+ (\cot\chi\cos\theta\sin\phi-\cos\phi  )\bm \partial_\theta+ (\cot\chi \csc\theta\cos\phi+\cot\theta \sin\phi )\bm \partial_\phi\nn \, ,\\
\bm R_3=&\cos\theta \,\bm \partial_\chi- \cot \chi\sin\theta \, \bm \partial_\theta -\bm \partial_\phi
\label{RIVF} \, .
\end{align}
One can verify that the two sets $\{\bm L_i\}$ and  $\{\bm R_i\}$ commute and that each of them close a $\mathfrak {su}(2)$ algebra (cf. \eqref{LRal})
\be
[\bm L_i,\bm R_j]=0,~~[\bm L_i,\bm L_j]=-2\epsilon_{ijk}\bm L_k,~~ ~~[\bm R_i,\bm R_j]=2\epsilon_{ijk}\bm R_k \, ,
\label{so4}
\ee
Moreover, $\{\bm L_i\}$ and $\{\bm R_i\}$ provide orthonormal basis in tangent space 
$$ \bm L_i\cdot\bm L_j =\delta_{ij},~~~~\bm R_i\cdot\bm R_j =\delta_{ij}$$
here $\bm A\cdot\bm B=\mathfrak g_{ab}A^aB^b$ denotes scalar product with \eqref{binv}\footnote{Similarly, the set of invariant forms satisfy 
$$ \bm \lambda_i\cdot\bm \lambda_j =\delta_{ij},~~~~\bm \rho_i\cdot\bm \rho_j =\delta_{ij}$$}.

It is also important to identify the $\mathfrak{so}(3)\subset\mathfrak{so}(4)$   which generates the SO$(3)$  stabiliser  in the coset representation $S^3 \simeq \tfrac{SO(4)}{SO(3)}$. This SO$(3)$ corresponds to  the isometries of  $S_{\theta,\phi}^2$ in \eqref{binv} with  associated Killing vectors   
\be
\begin{split}
\bm J_1=&  \sin \phi\, \bm \partial_\theta  +\cot\theta \cos\phi\,\bm \partial_\phi \, ,\\
\bm J_2=& - \cos\phi\, \bm \partial_\theta+ \cot\theta \sin\phi\, \bm \partial_\phi \, ,\\
\bm J_3=&- \bm \partial_\phi \, ,
\end{split}
\ee
These generators are related to ${\bm R}_i$, ${\bm L}_i$ as $\bm J_i=\frac12(\bm R_i-\bm L_i)$ (cf. \eqref{Js}). Defining the combination $  {\bm P}_i=\frac12(\bm R_i+\bm L_i)$ we find  
\begin{align}
{\bm P}_1=&\sin\theta \cos\phi \,\bm \partial_\chi+ \cot\chi\cos\theta\cos\phi\, \bm \partial_\theta -  \cot\chi \csc \theta\sin\phi \,\bm \partial_\phi\nn \, ,\\
{\bm P}_2=&  \sin\theta \sin\phi \,\bm \partial_\chi+ \cot\chi\cos\theta\sin\phi\, \bm \partial_\theta+\cot\chi \csc\theta\cos\phi\, \bm \partial_\phi \, ,
\label{ps}\\
{\bm P}_3=&\cos\theta \,\bm \partial_\chi- \cot \chi\sin\theta \, \bm \partial_\theta \, . \qquad\qquad\hspace{4cm}\nn 
\end{align}
The $\mathfrak{so}(4)$ algebra \eqref{so4} in the $\{\bm J_i,\bm P_i\}$-basis then takes the form
\begin{equation}
    [\bm J_i,\bm J_j]=\epsilon_{ijk}\bm J_k \, , \quad [{\bm P}_i, {\bm P}_j]= \epsilon_{ijk}\bm J_k\nn \, , \quad [{\bm J}_i, {\bm P}_j]= \epsilon_{ijk} {\bm P_k} \, ,
\end{equation}
Making $\bm P_i\mapsto\ell  \bm P_i $ and taking the $\ell\to\infty$ limit  we obtain the $\mathfrak{iso}(3)$ isometry algebra of $\mathbb R^3$ with  ${\bm J}_i$ and ${\bm P}_i$ the rotation and momentum generators. The Casimirs \eqref{casimir_gen} in the $\{J_i,P_i\}$ basis take the form
\begin{equation}
{\cal C}=\left[  (P_i)^2+(J_i)^2\right],~~~~\tilde{\cal C}=-   \,J\cdot P  
\end{equation}
where $J\cdot P=J_iP_i$. The second expression shows $\tilde {\cal C}$ plays the role of helicity of the irrep.

\nin {\sf   Maurer-Cartan identities} 
\be
 \bm d\bm\lambda+i\bm \lambda\wedge\bm \lambda=0~~\leadsto~~\bm d\bm\lambda^i-\epsilon^{ijk}\bm\lambda^j\wedge\bm\lambda^k=0
\label{MC1}
\ee
$$\bm d\bm\rho-i\bm \rho\wedge\bm \rho=0~~\leadsto~~\bm d\bm\rho^i+\epsilon^{ijk}\bm\rho^j\wedge\bm\rho^k=0$$
The dragging of the 1-forms $\bm \lambda^i$ and $\bm \rho^i$ along the Killing vectors $\bm L_i,\bm R_i$, using Cartan's formula   ${\cal L}_{\bm V}=i_{\bm V}\bm d+\bm di_{\bm V}$ and  \eqref{MC1}, gives 
\be 
{\cal L}_{\bm L_i}\bm\lambda^j
=-2\epsilon^{ijl}\bm\lambda^l,~~~~~~~{\cal L}_{\bm L_i}\bm\rho^j=0 \, ,
\ee
These results imply that $\bm\rho^i$ are invariant under  $\bm L_i$-dragging. Stated otherwise, the $\bm L_i$-vectors    generate right translations on $U$ (and viceversa for $\bm R_i$) \citep{gibbDAMTP, GS}. In terms of SO(4) representations the $[j_L,j_R]$ quantum numbers are
$$\bm L_i:~[0,1],~~~\qquad
\bm R_i:~[ 1,0]$$

\subsection*{Killing vectors and the Laplacian}
From the Killing vector equation
$$\nabla_a V_b+\nabla_b V_a=0~~\leadsto~~\nabla^a V_a=0$$
Contracting with $\nabla^a$, using \eqref{richo} and \eqref{s3g} one obtains\footnote{Our conventions for the Riemann tensor are $$[\nabla_a,\nabla_b]V^c=R^c{}_{dab}V^d$$}, 
\be
S^3\text{ \sf Killing vectors}:~~~- \nabla^2_{(1)}V_a=2 V_a 
\label{KVs}
\ee
This means Killing vectors of $S^3$ are transverse eigenfunctions of the Laplacian with eigenvalue 2 (cf. \eqref{vector_harmonics}).

\vspace{2mm}
\subsubsection*{Casimirs and Laplace-Beltrami operators}
The abstract Casimir operators in the SO(4) algebra can be translated into second order  differential operators acting on fields via the Lie derivative
\be 
{\cal C}=\frac12\big[(L_i)^2+(R_i)^2\big]~~\mapsto ~~ \frac12\big[\pounds_{\bm L_i}\pounds_{\bm L_i}+\pounds_{\bm R_i}\pounds_{\bm R_i}\big]
\label{CasiLie}
\ee
It is a well known fact that for maximally symmetric spaces this operators are related to the Laplace-Beltrami operator
$$\nabla^2=\mathfrak{g}^{ab}\nabla_a\nabla_b$$ 
with $\nabla_a$ the covariant derivative and  $a,b...$ denoting vector indices. In the following we quote the relations for scalar and vector fields.

\vspace{2mm}

\nin   {\sf Scalar fields}: the $S^3$-Laplace-Beltrami operator on scalars reads
\be
\nabla_{\mathfrak  (0)}^2 =\frac1{\sqrt {\mathfrak g}}\partial_a(\sqrt {\mathfrak g}\,\mathfrak g^{ab}\partial_b)=\left[\partial_\chi^2+2\cot\chi\partial_\chi+\frac1{\sin^2\chi}\nabla^2_{\!^{S^2}\!^{(0)}}\right] \, ,
\label{LB}
\ee
where
$$\nabla^2_{\!^{S^2}\!^{(0)}} =\left[\partial^2_\theta+ \cot\theta \partial_\theta +\frac1{\sin^2\theta}\partial_\phi^2\right]$$
The Lie derivative on scalars acts as
$$\pounds_{\bm V}\phi=\bm V[\phi]=V^a\partial_a \phi$$
The Casimir for the $\mathfrak{so}(3)$ subgalgebra generated by $\{\bm J_i\}$ coincides with the  Laplacian for $S^2_{\theta,\phi}$
\begin{equation}
     ( \pounds_{\bm J_i} )^2\phi= ({\bm J_i})^2\phi =\nabla^2_{\!^{S^2}}\phi \, ,
\end{equation}
Furthermore, the $\mathfrak{so}(4)$ Casimir $\cal C$ in \eqref{casimir_gen} acting on scalars coincides with the Laplace-Beltrami on $S^3$
\begin{equation}
 {\cal C}\phi=\frac{1}{2} \left[ ({\bm L_i})^2 +( {\bm R_i})^2\right]\phi = \nabla_{(0)}^2\phi \, ,
 \label{lb0}
\end{equation}
The remaining Casimir vanishes for scalars leading to\footnote{The vanishing of $\tilde{\cal C}$  follows from the spinless character of $\phi$.}
\be
\tilde{\cal C}\phi=-\bm J\cdot\bm P\phi=0~~\leadsto~~({\bm L_i})^2\phi=( {\bm R_i})^2\phi=\nabla_{(0)}^2 \phi
\label{ctil}
\ee
Expressing the non-trivial Casimir in the   $\{\bm J_i,\bm P_i\}$-basis one finds
\begin{equation}
\nabla_{(0)}^2 = ({\bm J_i})^2 +( {\bm P_i})^2  \, ,
\end{equation}
\subsection*{Scalar harmonics quantum numbers}
The eigenfunctions of the scalar Laplace-Beltrami operator \eqref{LB} are classified with quantum numbers $\tfrac k 2,l,m$ associated to $(\bm L_i)^2=(\bm R_i)^2$, $(\bm J_i)^2$ and $\bm J_3$ respectively. Explicitly\footnote{The  normalisation of the $\mathfrak{su}(2)$ generators in \eqref{so4} imply that the allowed values for the $L$-Casimir are  $-({\bm L_i})^2=4j_L(j_L+1)$ (similarly for the $R$-Casimir).},
$$  
 -\nabla_{(0)}^2{\sf Y}^{klm}=
   k(k+2) {\sf Y}^{klm} 
 $$
$$ -\nabla^2_{\!^{S^2}\!^{(0)}} {\sf Y}^{klm}  =
l(l+1) {\sf Y}^{klm}$$
\be
i\bm J_3 {\sf Y}^{klm}  =  m {\sf Y}^{klm} 
\label{scalar}
\ee
where  
$$k = 0,1,\cdots,~~~0\le l\le k~~~ \text{and}~~~   -l\le m\le l \, ,$$ 
From \eqref{lb0} and \eqref{ctil}
we conclude that scalar harmonics 
transform  under $  \mathfrak{su(2)}_L\times \mathfrak{su(2)}_R$ as
$$ {\sf Y}^{klm}:~ [ j_L,j_R]=[ \tfrac k2,\tfrac k2]$$
The explicit expression for the ${\sf Y}^{klm}$  can be obtained from the $S^2$ spherical harmonics $Y^{lm}$ inserting in \eqref{LB} the ansatz 
\begin{equation}
{\sf Y}^{klm}(\chi,\theta,\phi) =\frac1{\sqrt{a_{kl}}} R^{kl}(\chi) Y^{lm}(\theta,\phi) \, .
\label{scalar_harm}
\end{equation}
The solutions for $R^{kl}(\chi)$,  known as \emph{Fock} harmonics, take the form \citep{Gerlach:1978gy,Higuchi:1986wu,Klebanov:2011td,Lindblom:2017maa,Kumar:2020xjr},
\be 
R^{kl}(\chi) =  \sin^l\chi\,\frac{d^{l+1}(\cos (k+1)\chi)}{d(\cos \chi)^{l+1}},~~~  a_{kl} =\frac{(k+1)\pi}2\frac{(k+l+1)!}  {(k-l)!}  \,.
\ee
The orthogonality relations for ${\sf Y}^{klm}$ read 
\begin{equation}
\int d^3 \Omega \, {\sf Y}^{klm} ({\sf Y}^{k' l' m'})^* = \delta^{k k'} \delta^{l l'} \delta^{m m'}.
\end{equation}
The harmonics satisfy the following  properties under parity on $S^2$ and $S^3$
\begin{equation}
\begin{split}
\hat{P}_2({\sf Y}^{klm}) &\equiv {\sf Y}^{klm}(\chi, \pi-\theta,\pi+\phi) = (-1)^l {\sf Y}^{klm}(\chi,\theta,\phi) \\
\hat{P}_3({\sf Y}^{klm}) &\equiv {\sf Y}^{klm}(\pi-\chi, \pi-\theta,\pi+\phi) = (-1)^{k-1} {\sf Y}^{klm}(\chi,\theta,\phi)
\end{split}
\end{equation}

\subsection*{Vector fields and their quantum numbers}
The Lie derivative on vectors is
$$\pounds_{\bm V}\bm A=[\bm V,\bm A]= (V^a\partial_aA^b-A^a\partial_aV^b)\,\bm\partial_b$$
The relation \eqref{lb0} now turns into
\begin{equation}
{\cal C}A^a=\frac12\big[(\pounds_{\bm L_i})^2 +(\pounds_{\bm R_i})^2\big]A^a=\left(\nabla_{(1)}^2-2\right)A^a
\label{CasiA}
\end{equation}
whereas for $\tilde{\cal C}$, identified as an helicity operator in \citep{Higuchi:1991tn}, one obtains
\be
\tilde{\cal C} A^a=-\pounds_{\bm P_i}\pounds_{\bm J_i}A^a =\frac {\epsilon^{abc}}{\sqrt{\mathfrak g}}\partial_b A_c = \frac12  (\star  \bm{d  A})^a
\label{Casi2}
\ee
where $\epsilon^{123}=+1$ is the Levi-Civita tensor and $\star$ denotes de Hodge dual. The appearance of a ``mass''-looking term on the rhs of \eqref{CasiA} for non-zero spins is generically expected and arises from squaring the spinorial matrix $S_{ij}$ in the decomposition \eqref{Jdec}.

Traditionally, the vector harmonics are worked out from the scalar harmonics \eqref{scalar_harm} \citep{Higuchi:1986wu}. There are three different classes  denoted by $S_{(A)a}^{klm}$ with $A=0,1,2$. They satisfy
\begin{equation}
\begin{split}
-{ \nabla^2_{(1)}}  \bm S_{(0)}^{klm}&= \left(k(k+2) -2\right)\bm S_{(0) }^{klm} \, , \qquad   \nabla \cdot \bm S_{(0)}^{klm} =-\sqrt{k(k+2)}  {\sf Y}^{klm} ,\\
-{\nabla^2_{(1)}}  \bm S_{(1,2)}^{klm}&= \left(k(k+2) -1\right)\bm S_{(1,2)}^{klm} \, , \qquad   \nabla \cdot \bm S_{(1,2)}^{klm} = 0 \, , 
\end{split}
\label{vector_harmonics}
\end{equation}
$$-\nabla^2_{\!^{S^2}\!^{(1)}} \bm S_{(A)}^{klm} =l(l+1)\bm S_{(A)}^{klm},$$
$$i\pounds_{\bm J_3}\bm S_{(A)}^{klm}=m\bm S_{(A)}^{klm}$$
here $\nabla^2_{(1)}$ denotes the vector Laplacian on $S^3$. Notice $\bm S_{(0)}$ represent longitudinal modes while $\bm S_{(1,2)}$ correspond to transverse modes. In addition, comparing with \eqref{KVs} we learn that the $k=l=1$ eigenvectors coincide with the six Killing vectors\footnote{The equivalence of \eqref{vector_harmonics} and \eqref{KVs} was shown in \citep{Higuchi:1991tn}.} of $S^3$:
$$\{\bm L_i,\bm R_i\}\sim \{\bm S_{(1)}^{11m},\bm S_{(2)}^{11m}\},~~~~m=0,\pm1$$
Finally notice that the eigenvalues for the transverse modes in \eqref{vector_harmonics} match the general expression \eqref{eigenvalues} if we make $k\mapsto n$ and use $d=2$ for the 3-Sphere.

Vector harmonics are completely determined by the scalar ones \citep{Higuchi:1986wu,Lindblom:2017maa}. The longitudinal one is given by 
\be 
S^{klm}_{(0)a} = \frac{1}{\sqrt{k(k+2)}} \nabla_a {\sf Y}^{klm},~~~~k=1,2,..,~~~l=0,1,2,... 
\ee
while the transverse ones are given by\footnote{$\bm S^{klm}_{(1)}$ correspond to $\bm V^{(v;Llm)}$ and $\bm S^{klm}_{(2)}\leftrightarrow\bm V^{(s;Llm)}$ in the notation of \citep{Higuchi:1991tn}.}
\begin{align}
S^{klm}_{(1)a} &= \frac{1}{\sqrt{l(l+1)}} \varepsilon_a{}^{bc} \,\nabla_b {\sf Y}^{klm}\, \nabla_c \cos \chi\\
S^{klm}_{(2)a} &= \frac{1}{k+1}\varepsilon_a{}^{bc} \,\nabla_b S_{(1)c}^{klm}~\qquad~\qquad~~k,l\in\mathbb N,~ 1\le l\le k
\label{k>1}
\end{align}
here $\varepsilon^{abc}\equiv\epsilon^{abc}/\sqrt {\mathfrak{g}}$ is the covariant Levi-Civita tensor. Their orthogonality relations reads
\begin{equation}
\int d^3\Omega \sqrt{\mathfrak g} \, \mathfrak g^{ab} S^{klm}_{(A)a} (S^{  k' l' m'}_{(A')b})^*  = \delta_{AA'}\delta^{k k'} \delta^{l l'} \delta^{m m'} \, .
\end{equation}
While the completeness relation is 
\begin{equation}
    \sum_A\sum_{k,l,m} S_{(A)a}^{klm}(\Omega) \left( S_{(A)b}^{klm}(\bar{\Omega}) \right)^* = \frac{\delta^{(3)}(\Omega-\bar{\Omega})}{\sqrt{\mathfrak{g}}} \mathfrak{g}_{ab} \, , \quad A = 0,1,2
    \label{CRel}
\end{equation}
Thus, any vector field on $S^3$ admits an  expansion of the form 
\begin{equation}
\bm  V (\Omega) =\sum_{A} \sum_{k,l,m} V^{klm}_{(A)} \,\bm S^{klm}_{(A)}(\Omega) \, .
\end{equation}

We now discuss the SO(4) representation furnished by \eqref{vector_harmonics} in terms of $\mathfrak{su} (2)_L\times\mathfrak{su}(2)_R$ quantum numbers.  From \eqref{vector_harmonics} we recognise that $\bm S_{(0)}$ and $\bm S_{(1,2)}$ belong to different irreps as they have different ${\cal C}=\frac12\big[(\pounds_{\bm L_i})^2 +(\pounds_{\bm R_i})^2\big]$ eigenvalues
\begin{align} 
{\cal C}\bm S ^{klm}_{(0)} &= -k\left(k+2 \right) \bm {S}^{klm}_{(0)} \nn\\
{\cal C}\bm S ^{klm}_{(1,2)}  & = -\left( k+1 \right)^2 \bm{S}^{klm}_{(1,2) }
\end{align}
It turns out that we need to combine $\bm S_{(1,2)}$ to obtain $\tilde{\cal C}$ eigenstates. Defining\footnote{Notice there is no relative $i$ factor in \eqref{qVec} as compared to the case of circular polarizations in flat space. This implies that under conjugation $(\bm {\mathcal  S}_{(\pm)})^\dagger=(-)^m\bm{\mathcal  S}^{kl(-m)}_{(\pm)}$.}  
\begin{equation}
\bm{\mathcal{S}}^{klm}_{(\pm)} = \frac{1}{\sqrt{2}} (\bm S^{klm}_{(2)} \pm \bm S^{klm}_{(1) }) \, , \qquad \bm{\mathcal{S}}^{klm}_{(0) } = \bm S^{klm}_{(0) } \, ,
\label{qVec}
\end{equation}
we find for $\tilde{\cal C}=\frac12\big[(\pounds_{\bm L_i})^2 -(\pounds_{\bm R_i})^2\big] $ that,
$$\tilde{\mathcal{C}} \bm{\mathcal{S}}^{klm}_{(0)} = 0\,,$$
\begin{equation}
\tilde{\mathcal{C}} \bm{\mathcal{S}}^{klm}_{(\pm) } = \mp (k+1) \bm{\mathcal{S}}^{klm}_{(\pm) }  \, , 
\label{qtilVec}
\end{equation}
Comparing \eqref{qVec},\eqref{qtilVec} with \eqref{su2x} we find 
$\left[ j_L, j_R\right]$ to be 
\begin{equation}
\bm{\mathcal{S}}^{klm}_{(0)} : \left[ \frac{k}{2}, \frac{k}{2} \right] \, , \qquad \bm{\mathcal{S}}^{klm}_{(\pm)}: \left[  \frac{k\pm 1}{2}, \frac{k\mp 1}{2} \right] \,~~~\text{with}~~k=1,2,.. .
\label{Ss}
\end{equation} 
Recalling that $\tilde{\mathcal{C}}$ gives the projection of  ${\bm J_i} $ on the direction of ${\bm P_i}$,  the modes $\bm{\mathcal{S}}_{(\pm) }^{klm}$ have opposite Chirality. The result of vectors transforming in $j_L=j_R$, $j_L=j_R+1$ and $j_R=j_L+1$ UIRs of SO(4) can be understood as follows. Any vector field can be expanded  as a linear combination of left-invariant vectors $\bm L_i$. These LIV transform in the [0,1] irrep. The coefficients of an arbitrary vector  in the $\{\bm L_i\}$-basis can in turn be expanded in terms of ${\sf Y}^{klm}$, which as mentioned above transform in the $[\tfrac k2, \tfrac k2]$. Hence, we conclude that
$$\bigoplus_k\left([1,0]\otimes[\tfrac k2, \tfrac k2]\right)=\bigoplus_k \big( \underbrace{[\tfrac k2-1,\tfrac k2]}_{{\cal S}_{(-)}}\oplus\underbrace{[\tfrac k2 ,\tfrac k2]}_{{\cal S}_{(0)}}\oplus\underbrace{[\tfrac k2+1,\tfrac k2]}_{{\cal S}_{(+)}} \big)$$
here $k=0,1,2,...$. Notice that when $k = 0$ the first two terms are absent and when $k=1$ first term in the rhs is absent. In this way we recover \eqref{Ss}.

\vspace{2mm}

\nin{\sf A word on notation}. SO(4) is rank-2 and not simple, hence it is customary to find two alternative notations to label its irreps: 

(i) $[j_L,j_R]$: in terms of $  \mathfrak{su(2)}_L\times \mathfrak{su(2)}_R$ quantum numbers,

(ii) $\bm s=(s_1,\pm s_2)$ ($ \mathbb{Y}_{s_1,s_2}$): in terms of the two Cartan of SO(4). 

\nin Being rank-2, Young tableaux consist of two rows. Scalar harmonics on $S^d$ correspond to single-row representations with $n$-boxes ($\bm s=(n,0)$),  these correspond to $\mathbb{Y}_n $ of SO($d + 1$) in the text. On the other hand, transverse vector harmonics correspond to two-row representations with one box in the second row, these are $\mathbb{Y}_{\bm s}=\mathbb{Y}_{n,\pm1}$  in the text. 
The $\pm$ notation in the irreps relate to the sign  of the second Casimir, namely $\tilde{\mathcal{C}} $. The relation between the two notations is
$$s_1=j_L+j_R,~~s_2=j_L-j_R~\leftrightsquigarrow~j_L=\tfrac12(s_1+s_2),~j_R=\tfrac12(s_1-s_2)$$

\section{de Sitter geometry}
\label{kills}

The $D=d+1$ dimensional de Sitter space is a maximally symmetric spacetime with
$$R_{\mu\nu\alpha\beta}[g]=\frac1{\ell^2}(g_{\mu\alpha}g_{\nu\beta}-g_{\nu\alpha}g_{\mu\beta}),~~~R_{\mu\nu}[g]=\frac1{\ell^2}(D-1)g_{\mu\nu},~~~R[g]=\frac{D(D-1)}{\ell^2}\,.$$
We will denote de Sitter space alternatively as $dS_D$ or $  dS_{d+1}$.\\
The hyperboloid (\ref{def:emb_mink}) can be fully covered with  
\begin{equation}
X^0 =\ell\, \cot t \, , \qquad X^i =\ell\, \frac1{\sin t}\, n^i \, ,~~~~i=1,..,d+1
\label{def:global_emb}
\end{equation}
here $t\in(-\pi,0)$, and $n^i\in\mathbb R^{d+1}$ is a unit vector.  The induced metric on the hyperboloid is  
\begin{equation}
\text{\sf de Sitter}_{D}:~~~\frac{ds^2}{\ell^2} = \frac{-dt^2 +  d\Omega_d^2}{\sin^2t}  , 
\label{dSd+1}
\end{equation}
where   $d\Omega_d^2$ denotes the round metric on $S^d$.  Notice that the radius of  $S^d$ shrinks for $t\in(-\pi,-\pi/2)$ and  expands for $t\in(-\pi/2,0)$. 

\vspace{3mm}

\nin{\sf de Sitter$_4$}: we now consider the case $D=4$. The isometry group is SO$(1,4)$, and we parameterise the unit vector $n_i$ in \eqref{def:global_emb} as 
$$n^1=\sin\chi\sin\theta\cos\phi,~~~n^2=\sin\chi\sin\theta\sin\phi,~~~n^3=\sin\chi\cos\theta,~~~n^4=\cos\chi$$
With this parametrisation the round 3-sphere metric reads
$$d\Omega_3^2=d\chi^2+\sin^2\chi(d\theta^2+\sin^2\theta d\phi^2)$$
. {\sf Christoffell symbols}:
\be
\Gamma^t_{tt}=\Gamma^t_{\chi\chi}=\Gamma^\chi_{t\chi}=\Gamma^\theta_{t\theta}=\Gamma^\phi_{t\phi}=\frac{\Gamma^t_{\theta\theta}}{\sin^2\chi}=\frac{\Gamma^t_{\phi\phi}}{\sin^2\chi\sin^2\theta}=-\cot t
\label{ChdS}
\ee
$$ \Gamma^\chi_{\theta\theta}=\frac{\Gamma^\chi_{\phi\phi}}{\sin^2\theta}=-\cos\chi\sin\chi,~~~ \Gamma^\theta_{\chi\theta }=\Gamma^\phi_{\chi\phi }=\cot\chi, ~~\Gamma^\theta_{\phi\phi}=-\cos\theta\sin\theta,~~ \Gamma^\phi_{\theta\phi}=\cot\theta $$

\nin . {\sf Killing vectors}: 
\be 
\bm D = \sin t \cos\chi \,\bm \partial_t +\cos t \sin \chi\, \bm \partial_\chi
\label{l0}
\ee
\begin{align}
\bm J_1 &=  \sin\phi \,\bm\partial_\theta + \cot \theta\cos \phi  \,\bm\partial_\phi\nn\\  
\bm J_2 &=-\cos\phi \,\bm \partial_\theta + \cot \theta\sin \phi  \,\bm \partial_\phi\nn\\    
\bm J_3 &= -\bm\partial_\phi
\label{Js}
\end{align} 
\begin{align}
\bm M_{\pm 1} =& \mp \sin t\sin\chi \sin \theta \cos\phi \,\bm\partial_t  + (1\pm \cos t \cos\chi) \sin\theta \cos\phi \,\bm\partial_\chi \nn\\ 
&  + (\cot\chi \pm \cos t \csc\chi)(\cos\theta \cos\phi \,\bm \partial_\theta - \csc\theta \sin\phi \,\bm \partial_\phi) \nn  \\ 
\bm M_{\pm 2} = &\mp \sin t\sin \chi \sin \theta \sin\phi \,\bm  \partial_t  + (1\pm \cos t\cos\chi) \sin\theta \sin\phi \,\bm \partial_\chi \nn\\ 
&    + (\cot \chi \pm \cos t \csc \chi)(\cos \theta \sin \phi \,\bm \partial_\theta + \csc\theta \cos\phi\, \bm \partial_\phi) \, \nn \\ 
\bm M_{\pm 3}  = &\mp \sin t \sin \chi \cos\theta\, \bm\partial_t  + (1\pm \cos t \cos\chi) \cos\theta\,\bm\partial_\psi\nn\\
&- (\cot\chi \pm \cos t \csc \chi) \sin\theta\,\bm \partial_\theta  
\label{Ms}
\end{align} 
Their algebra is
$$[\bm J_i,\bm J_j]=\epsilon_{ijk}\,\bm J_k,~~~~[\bm J_i,\bm M_{\pm j}]=\epsilon_{ijk}\bm M_{\pm k}$$
\be
[\bm D,\bm M_{\pm i}]=\mp\bm M_{\pm i},~~~~[\bm M_{\pm i},\bm M_{\mp j}]=2\bm D\,\delta_{ij}+2\epsilon_{ijk}\,\bm J_k
\label{dsAl}
\ee
These commutation relations define the SO(1,4) algebra. The (quadratic) Casimir ${\cal C}_{\text{SO}(1,4)}$ of the algebra is
\be
{\cal C}_{\text{SO}(1,4)}=-\bm D(\bm D-3)+\bm M_{-i}\bm M_{+i}+\bm J^2
\label{Casds}
\ee
As expected from the previous section,  this Casimir acting as Lie derivatives (see \eqref{CasiLie}) relates to the Laplace-Beltrami operator on dS. In particular
$$\text{\sf Casimir on scalars}:~~~{\cal C}_{\text{SO}(1,4)}\phi=g^{\mu\nu}D_\mu D_\nu\phi$$
whereas for vectors we have
\be
\text{\sf Casimir on vectors}:~~~{\cal C}_{\text{SO}(1,4)}A^\mu=(g^{\mu\nu}D_\mu D_\nu-3)A^\mu
\label{Cvec}
\ee
here $D_\mu$ is the covariant derivative on de Sitter spacetime.

The maximal compact subgroup SO(4)$\,\subset\,$SO(4,1) acts closely on $S^3$. Their generators are given by the $\{\bm J_i,\bm P_i\}$ where (cf. \eqref{ps})
$$\bm P_i=\tfrac12(\bm M_{+i}+\bm M_{-i})~\leadsto~\left\{\begin{array}{l}
\bm P_1=   \sin\theta \cos\phi \,\bm \partial_\chi+  \cot \chi  \cos \theta \cos \phi\, \bm \partial_\theta -\cot \chi  \csc\theta \sin\phi \,\bm \partial_\phi\\
\bm P_2=   \sin\theta \sin\phi \,\bm \partial_\chi+  \cot \chi  \cos \theta \sin \phi\, \bm \partial_\theta +\cot \chi  \csc\theta \cos\phi \,\bm \partial_\phi\\
{\bm P}_3=\cos\theta \,\bm \partial_\chi- \cot \chi\sin\theta \, \bm \partial_\theta \, . 
\end{array}
\right.
$$
In a similar way we define the boost operators 
\begin{equation}
 \bm K_i=\tfrac12(\bm M_{+i}-\bm M_{-i})
\end{equation}
which give
\begin{align} 
\bm K_1= & -\sin t\sin\chi \sin\theta \cos\phi \,\bm\partial_t +\cos t\cos\chi\sin\theta\cos\phi \,\bm \partial_\chi\nn\\
&+ \cos t \csc \chi  \cos \theta \cos \phi\, \bm \partial_\theta -\cos t\csc \chi  \csc\theta \sin\phi \,\bm \partial_\phi\\
\bm K_2= &  -\sin t\sin\chi \sin\theta \sin\phi \,\bm\partial_t +\cos t\cos\chi\sin\theta\sin\phi \,\bm \partial_\chi\nn\\
&+ \cos t \csc \chi  \cos \theta \sin \phi\, \bm \partial_\theta +\cos t\csc \chi  \csc\theta \cos\phi \,\bm \partial_\phi\\
{\bm K}_3=&-\sin t\sin\chi\cos\theta\,\bm\partial_t+\cos t\cos\chi\cos\theta \,\bm \partial_\chi- \cos t\csc \chi\sin\theta \, \bm \partial_\theta \, . 
\end{align}
For completeness we write the algebra in the $\{\bm D,\bm J_i,\bm P_i,\bm K_i\}$ basis. It reads
$$[\bm J_i,\bm J_j]=\epsilon_{ijk}\,\bm J_k,~~~~[\bm J_i,\bm P_j]=\epsilon_{ijk}\bm P_k,~~~~[\bm J_i,\bm K_j]=\epsilon_{ijk}\bm K_k$$
$$[\bm P_i,\bm P_j]=\epsilon_{ijk}\,\bm J_k,~~~~[\bm K_i,\bm K_j]= \epsilon_{ijk}\bm J_k$$
$$[\bm D,\bm P_i]=-\bm K_i,~~~~[\bm D,\bm K_i]=-\bm P_i,~~~~[\bm P_i ,\bm K_j]=-\bm D\,\delta_{ij}$$
The Casimir \eqref{Casds} takes the form
$${\cal C}_{\text{SO}(1,4)}= -\bm D^2-(\bm K_i)^2+ (\bm P_i)^2+(\bm J_i)^2 $$

A couple of comments are in order:\\
i. Under the identification $\bm D\mapsto   D$, $\bm M_{+i}\mapsto  K_i$, $\bm M_{-i}\mapsto P_i$ the algebra \eqref{dsAl} coincides with \eqref{def:conf_algebra}. It is important to remark that  although  $\bm M_{\pm i}$ act as ladders wrt $\bm D$, the reality conditions in de Sitter differ from those implemented in traditional CFT, here $$(\bm M_{\pm i})^\dagger=-\bm M_{\pm i}$$ 
which differs from  $(\bm M_{\pm i})^\dagger=-\bm M_{\mp i}$ (equivalently to $ (P_i)^\dagger=-K_i$) appearing in Lorentzian CFT.\\
ii. If we evaluate the de Sitter Killing vectors \eqref{l0}-\eqref{Ms}  on future infinity ${\cal I}^+$, i.e. $t=0$ we obtain the conformal Killing vectors of $S^3$ with non-zero divergence.\\
iii. It is amusing to notice that the conformal Killing vectors on $S^3$ referred to in ii. are closed,  they can be derived from potentials. Explicitly, in the asymptotic future $t=0$,
\begin{align}
(\bm D)_\mu&=\partial_\mu(-\cos\chi)\nn\\
(\bm K_1)_\mu&=\partial_\mu(\sin\chi\,\sin\theta\,\cos\phi)\nn\\
(\bm K_2)_\mu&=\partial_\mu(\sin\chi\,\sin\theta\,\sin\phi)\nn\\
(\bm K_3)_\mu&=\partial_\mu(\sin\chi\,\cos\theta )\nn
\end{align}

\nin . {\sf Conserved inner product and de Sitter invariance}: given two eigenfunctions $A_\mu^{(1,2)}$ of the Laplace-Beltrami operator possesing the same eigenvalue, namely
\be
\square_{dS}A^{(1)}_\mu=\lambda A^{(1)}_\mu,~~\square_{dS}A^{(2)}_\mu=\lambda A^{(2)}_\mu\,,
\label{boxes}
\ee
an inner product for gauge field configurations is constructed out from the   current
\be
j_\mu[A^{(1)},A^{(2)}]:= i\left[(A_\rho^{(1)})^*D_\mu A^{ (2)\,\rho} -(A_\rho^{(2)})^*D_\mu A^{ (1)\,\rho}\right]~\leadsto~D_\mu j^\mu=0\,.
\label{jmu}
\ee
Its conservation follows from \eqref{boxes}. The expression for the inner product in global coordinates is
\begin{align}
( A^{(1)},A^{(2)})&:= -i\int d\Omega\sqrt{-g}j^t
\label{InnPr}
\end{align}
To show the invariance of the inner product under de Sitter isometries we compute the variation $\delta j^\mu$ under the action of a de Sitter Killing vector $K^\mu$
$$\delta j^\mu=\pounds_{\bm K}j^\mu
=K^\rho D_\rho j^\mu-j^\rho D_\rho K^\mu=D_\rho(K^\rho  j^\mu-j^\rho  K^\mu) $$
in passing to the last expression on the right we used that $\bm j$ and $\bm K$ are conserved. It then follows that the integrand in \eqref{InnPr} changes as
$$\sqrt{-g}\,\delta j^t=\partial_a(\sqrt{-g}(K^aj^t-K^tj^a))\,.$$
Being a total derivative, it vanishes upon integration over the closed spatial sections of de Sitter.
 
\section{Glossary}
\begin{itemize}
    \item{$L_{AB}$: anti-hermitian generators of SO$(1,d+1)$.}
    \item{$K_{ij}$: anti-hermitian generators of SO$(d+1)$.}
    \item{$g_{\mu \nu}$: Spacetime metric.}
    \item{$\mathfrak{g}_{ab}$: $3-$sphere metric.}
    \item{$\bm{L}_i,\bm{R}_i$: Generators of SO$(4)$ under the SO$(4)\simeq$SU$(2)_L\times$SU$(2)_R$ decomposition.}
    \item{$\bm{J}_i$: Generators of SO$(3)$ in the coset representation $S^3 \simeq \frac{SO(4)}{SO(3)}$.}
    \item{$\bm{D}, \bm{K}_i$: de Sitter boosts}
\end{itemize}

\bibliographystyle{utphys2}
\bibliography{refs.bib}

\end{document}